\documentclass[journal,onecolumn,draftcls,11pt]{IEEEtran}

\usepackage{amssymb, amsmath, epsfig}
\usepackage{amsfonts}
\usepackage{bbm}
\usepackage{mathrsfs}
\usepackage{xspace}
\usepackage{bm}

\usepackage{cite}

\newtheorem{theorem}{Theorem}

\newtheorem{proposition}{Prop.}
\newtheorem{lemma}{Lemma}

\newtheorem{assumption}{Assumption}

\newenvironment{textbmatrix}{   \setlength{\arraycolsep}{2.5pt}%
                                                                \big[\begin{matrix}}{\end{matrix}\big]%
                                                                \raisebox{0.08ex}{\vphantom{M}}}

\def\be{\begin{equation}}
\def\ee{\end{equation}}
\def\een{\nonumber \end{equation}}
\def\mat{\begin{bmatrix}}
\def\emat{\end{bmatrix}}
\def\btm{\begin{textbmatrix}}
\def\etm{\end{textbmatrix}}

\def\ba#1\ea{\begin{align}#1\end{align}}
\def\bs#1\es{\begin{split}#1\end{split}}
\def\bg#1\eg{\begin{gather}#1\end{gather}}
\def\bi#1\ei{\begin{itemize}#1\end{itemize}}

\newcommand{\safemath}[2]{\newcommand{#1}{\ensuremath{#2}\xspace}}

\DeclareMathOperator{\Tr}{Tr}                           
\DeclareMathOperator{\diag}{diag}                       
\DeclareMathOperator*{\argmax}{arg\;max}                
\DeclareMathOperator{\Exop}{\mathbb{E}}         
\DeclareMathOperator{\Varop}{\mathbb{V}\!\mathrm{ar}} 

\safemath{\interior}{\mathrm{Int}}                       
\newcommand{\tp}[1]{\ensuremath{#1^{T}}}                
\newcommand{\herm}[1]{\ensuremath{#1^{H}}}      

\safemath{\dfn}{:=}                                                     
\safemath{\dirac}{\delta}                                       

\safemath{\SNR}{\text{\sc snr}}                                 
\safemath{\No}{N_0}                                                     
\safemath{\Es}{E_s}                                                     
\safemath{\Eb}{E_b}                                                     
\safemath{\EbNo}{\frac{\Eb}{\No}}
\safemath{\EsNo}{\frac{\Es}{\No}}

\DeclareMathOperator{\CHop}{\ensuremath{\mathbb{H}}} 
\safemath{\tvir}{h_{\CHop}}                                     
\safemath{\tvtf}{L_{\CHop}}                                     
\safemath{\spf}{S_{\CHop}}                                              
\safemath{\bff}{H_{\CHop}}                                      

\safemath{\ircf}{R_{h}}                                         
\safemath{\scf}{R_{S}}                                          
\safemath{\tfcf}{R_{L}}                                         
\safemath{\bfcf}{R_{H}}                                         

\safemath{\mi}{I}                                                       
\safemath{\capacity}{C}                                         

\safemath{\normal}{\mathcal{N}}                         
\safemath{\circnorm}{\mathcal{CN}}                      
\safemath{\mchain}{\leftrightarrow}                     

\safemath{\dB}{\,\mathrm{dB}}
\safemath{\dBm}{\,\mathrm{dBm}}
\safemath{\Hz}{\,\mathrm{Hz}}
\safemath{\kHz}{\,\mathrm{kHz}}
\safemath{\MHz}{\,\mathrm{MHz}}
\safemath{\GHz}{\,\mathrm{GHz}}
\safemath{\s}{\,\mathrm{s}}
\safemath{\ms}{\,\mathrm{ms}}
\safemath{\mus}{\,\mathrm{\mu s}}
\safemath{\ns}{\,\mathrm{ns}}
\safemath{\meter}{\,\mathrm{m}}
\safemath{\mm}{\,\mathrm{mm}}
\safemath{\cm}{\,\mathrm{cm}}
\safemath{\m}{\,\mathrm{m}}
\safemath{\W}{\,\mathrm{W}}
\safemath{\J}{\,\mathrm{J}}
\safemath{\K}{\,\mathrm{K}}
\safemath{\bit}{\,\mathrm{bit}}

\safemath{\define}{\triangleq}                  

\safemath{\equivalent}{\sim}
\safemath{\distas}{\sim}                                        

\safemath{\reals}{\mathbb{R}}
\safemath{\positivereals}{\mathbb{R}^{+}}
\safemath{\integers}{\mathbb{Z}}
\safemath{\posint}{\mathbb{Z}_{+}}
\safemath{\naturals}{\mathbb{N}}
\safemath{\complexset}{\mathbb{C}}

\safemath{\setA}{\mathcal{A}}
\safemath{\setB}{\mathcal{B}}
\safemath{\setC}{\mathcal{C}}
\safemath{\setD}{\mathcal{D}}
\safemath{\setE}{\mathcal{E}}
\safemath{\setF}{\mathcal{F}}
\safemath{\setG}{\mathcal{G}}
\safemath{\setH}{\mathcal{H}}
\safemath{\setI}{\mathcal{I}}
\safemath{\setJ}{\mathcal{J}}
\safemath{\setK}{\mathcal{K}}
\safemath{\setL}{\mathcal{L}}
\safemath{\setM}{\mathcal{M}}
\safemath{\setN}{\mathcal{N}}
\safemath{\setO}{\mathcal{O}}
\safemath{\setP}{\mathcal{P}}
\safemath{\setQ}{\mathcal{Q}}
\safemath{\setR}{\mathcal{R}}
\safemath{\setS}{\mathcal{S}}
\safemath{\setT}{\mathcal{T}}
\safemath{\setU}{\mathcal{U}}
\safemath{\setV}{\mathcal{V}}
\safemath{\setW}{\mathcal{W}}
\safemath{\setX}{\mathcal{X}}
\safemath{\setY}{\mathcal{Y}}
\safemath{\setZ}{\mathcal{Z}}
\safemath{\emptySet}{\varnothing}

\safemath{\bma}{\mathbf{a}}
\safemath{\bmb}{\mathbf{b}}
\safemath{\bmc}{\mathbf{c}}
\safemath{\bmd}{\mathbf{d}}
\safemath{\bme}{\mathbf{e}}
\safemath{\bmf}{\mathbf{f}}
\safemath{\bmg}{\mathbf{g}}
\safemath{\bmh}{\mathbf{h}}
\safemath{\bmi}{\mathbf{i}}
\safemath{\bmj}{\mathbf{j}}
\safemath{\bmk}{\mathbf{k}}
\safemath{\bml}{\mathbf{l}}
\safemath{\bmm}{\mathbf{m}}
\safemath{\bmn}{\mathbf{n}}
\safemath{\bmo}{\mathbf{o}}
\safemath{\bmp}{\mathbf{p}}
\safemath{\bmq}{\mathbf{q}}
\safemath{\bmr}{\mathbf{r}}
\safemath{\bms}{\mathbf{s}}
\safemath{\bmt}{\mathbf{t}}
\safemath{\bmu}{\mathbf{u}}
\safemath{\bmv}{\mathbf{v}}
\safemath{\bmw}{\mathbf{w}}
\safemath{\bmx}{\mathbf{x}}
\safemath{\bmy}{\mathbf{y}}
\safemath{\bmz}{\mathbf{z}}

\bmdefine{\biad}{a}
\bmdefine{\bibd}{b}
\bmdefine{\bicd}{c}
\bmdefine{\bidd}{d}
\bmdefine{\bied}{e}
\bmdefine{\bifd}{f}
\bmdefine{\bigd}{g}
\bmdefine{\bihd}{h}
\bmdefine{\biid}{i}
\bmdefine{\bijd}{j}
\bmdefine{\bikd}{k}
\bmdefine{\bild}{l}
\bmdefine{\bimd}{m}
\bmdefine{\bind}{n}
\bmdefine{\biod}{o}
\bmdefine{\bipd}{p}
\bmdefine{\biqd}{q}
\bmdefine{\bird}{r}
\bmdefine{\bisd}{s}
\bmdefine{\bitd}{t}
\bmdefine{\biud}{u}
\bmdefine{\bivd}{v}
\bmdefine{\biwd}{w}
\bmdefine{\bixd}{x}
\bmdefine{\biyd}{y}
\bmdefine{\bizd}{z}

\bmdefine{\bixid}{\xi}
\bmdefine{\bilambdad}{\lambda}
\bmdefine{\bimud}{\mu}
\bmdefine{\bithetad}{\theta}
\bmdefine{\biphid}{\phi}

\safemath{\bmia}{\biad}
\safemath{\bmib}{\bibd}
\safemath{\bmic}{\bicd}
\safemath{\bmid}{\bidd}
\safemath{\bmie}{\bied}
\safemath{\bmif}{\bifd}
\safemath{\bmig}{\bigd}
\safemath{\bmih}{\bihd}
\safemath{\bmii}{\biid}
\safemath{\bmij}{\bijd}
\safemath{\bmik}{\bikd}
\safemath{\bmil}{\bild}
\safemath{\bmim}{\bimd}
\safemath{\bmin}{\bind}
\safemath{\bmio}{\biod}
\safemath{\bmip}{\bipd}
\safemath{\bmiq}{\biqd}
\safemath{\bmir}{\bird}
\safemath{\bmis}{\bisd}
\safemath{\bmit}{\bitd}
\safemath{\bmiu}{\biud}
\safemath{\bmiv}{\bivd}
\safemath{\bmiw}{\biwd}
\safemath{\bmix}{\bixd}
\safemath{\bmiy}{\biyd}
\safemath{\bmiz}{\bizd}

\safemath{\bmxi}{\bixid}
\safemath{\bmlambda}{\bilambdad}
\safemath{\bmmu}{\bimud}
\safemath{\bmtheta}{\bithetad}
\safemath{\bmphi}{\biphid}

\safemath{\bA}{\mathbf{A}}
\safemath{\bB}{\mathbf{B}}
\safemath{\bC}{\mathbf{C}}
\safemath{\bD}{\mathbf{D}}
\safemath{\bE}{\mathbf{E}}
\safemath{\bF}{\mathbf{F}}
\safemath{\bG}{\mathbf{G}}
\safemath{\bH}{\mathbf{H}}
\safemath{\bI}{\mathbf{I}}
\safemath{\bJ}{\mathbf{J}}
\safemath{\bK}{\mathbf{K}}
\safemath{\bL}{\mathbf{L}}
\safemath{\bM}{\mathbf{M}}
\safemath{\bN}{\mathbf{N}}
\safemath{\bO}{\mathbf{O}}
\safemath{\bP}{\mathbf{P}}
\safemath{\bQ}{\mathbf{Q}}
\safemath{\bR}{\mathbf{R}}
\safemath{\bS}{\mathbf{S}}
\safemath{\bT}{\mathbf{T}}
\safemath{\bU}{\mathbf{U}}
\safemath{\bV}{\mathbf{V}}
\safemath{\bW}{\mathbf{W}}
\safemath{\bX}{\mathbf{X}}
\safemath{\bY}{\mathbf{Y}}
\safemath{\bZ}{\mathbf{Z}}

\safemath{\bZero}{\mathbf{0}}

\bmdefine{\biAd}{A}
\bmdefine{\biBd}{B}
\bmdefine{\biCd}{C}
\bmdefine{\biDd}{D}
\bmdefine{\biEd}{E}
\bmdefine{\biFd}{F}
\bmdefine{\biGd}{G}
\bmdefine{\biHd}{H}
\bmdefine{\biId}{I}
\bmdefine{\biJd}{J}
\bmdefine{\biKd}{K}
\bmdefine{\biLd}{L}
\bmdefine{\biMd}{M}
\bmdefine{\biOd}{N}
\bmdefine{\biPd}{O}
\bmdefine{\biQd}{P}
\bmdefine{\biRd}{R}
\bmdefine{\biSd}{S}
\bmdefine{\biTd}{T}
\bmdefine{\biUd}{U}
\bmdefine{\biVd}{V}
\bmdefine{\biWd}{W}
\bmdefine{\biXd}{X}
\bmdefine{\biYd}{Y}
\bmdefine{\biZd}{Z}

\bmdefine{\biDelta}{\Delta}
\bmdefine{\biLambda}{\Lambda}
\bmdefine{\biPhi}{\Phi}
\bmdefine{\biSigma}{\Sigma}
\bmdefine{\biOmega}{\Omega}
\bmdefine{\biTheta}{\Theta}

\safemath{\bimA}{\biAd}
\safemath{\bimB}{\biBd}
\safemath{\bimC}{\biCd}
\safemath{\bimD}{\biDd}
\safemath{\bimE}{\biEd}
\safemath{\bimF}{\biFd}
\safemath{\bimG}{\biGd}
\safemath{\bimH}{\biHd}
\safemath{\bimI}{\biId}
\safemath{\bimJ}{\biJd}
\safemath{\bimK}{\biKd}
\safemath{\bimL}{\biLd}
\safemath{\bimM}{\biMd}
\safemath{\bimN}{\biNd}
\safemath{\bimO}{\biOd}
\safemath{\bimP}{\biPd}
\safemath{\bimQ}{\biQd}
\safemath{\bimR}{\biRd}
\safemath{\bimS}{\biSd}
\safemath{\bimT}{\biTd}
\safemath{\bimU}{\biUd}
\safemath{\bimV}{\biVd}
\safemath{\bimW}{\biWd}
\safemath{\bimX}{\biXd}
\safemath{\bimY}{\biYd}
\safemath{\bimZ}{\biZd}

\safemath{\bDelta}{\bielta}
\safemath{\bLambda}{\biLambda}
\safemath{\bPhi}{\biPhi}
\safemath{\bSigma}{\biSigma}
\safemath{\bOmega}{\biOmega}
\safemath{\bTheta}{\biTheta}

\safemath{\veca}{\bma}
\safemath{\vecb}{\bmb}
\safemath{\vecc}{\bmc}
\safemath{\vecd}{\bmd}
\safemath{\vece}{\bme}
\safemath{\vecf}{\bmf}
\safemath{\vecg}{\bmg}
\safemath{\vech}{\bmh}
\safemath{\veci}{\bmi}
\safemath{\vecj}{\bmj}
\safemath{\veck}{\bmk}
\safemath{\vecl}{\bml}
\safemath{\vecm}{\bmm}
\safemath{\vecn}{\bmn}
\safemath{\veco}{\bmo}
\safemath{\vecp}{\bmp}
\safemath{\vecq}{\bmq}
\safemath{\vecr}{\bmr}
\safemath{\vecs}{\bms}
\safemath{\vect}{\bmt}
\safemath{\vecu}{\bmu}
\safemath{\vecv}{\bmv}
\safemath{\vecw}{\bmw}
\safemath{\vecx}{\bmx}
\safemath{\vecy}{\bmy}
\safemath{\vecz}{\bmz}
\safemath{\vecZero}{\bZero}

\safemath{\vecxi}{\bmxi}
\safemath{\veclambda}{\bmlambda}
\safemath{\vecmu}{\bmmu}
\safemath{\vectheta}{\bmtheta}
\safemath{\vecphi}{\bmphi}

\safemath{\matA}{\bA}
\safemath{\matB}{\bB}
\safemath{\matC}{\bC}
\safemath{\matD}{\bD}
\safemath{\matE}{\bE}
\safemath{\matF}{\bF}
\safemath{\matG}{\bG}
\safemath{\matH}{\bH}
\safemath{\matI}{\bI}
\safemath{\matJ}{\bJ}
\safemath{\matK}{\bK}
\safemath{\matL}{\bL}
\safemath{\matM}{\bM}
\safemath{\matN}{\bN}
\safemath{\matO}{\bO}
\safemath{\matP}{\bP}
\safemath{\matQ}{\bQ}
\safemath{\matR}{\bR}
\safemath{\matS}{\bS}
\safemath{\matT}{\bT}
\safemath{\matU}{\bU}
\safemath{\matV}{\bV}
\safemath{\matW}{\bW}
\safemath{\matX}{\bX}
\safemath{\matY}{\bY}
\safemath{\matZ}{\bZ}
\safemath{\matZero}{\bZero}

\safemath{\matDelta}{\bDelta}
\safemath{\matLambda}{\bLambda}
\safemath{\matPhi}{\bPhi}
\safemath{\matSigma}{\bSigma}
\safemath{\matOmega}{\bOmega}
\safemath{\matTheta}{\bTheta}

\safemath{\matIdentity}{\matI}

\newcommand{\utility[2]}{u_{#1}(#2)}
\newcommand{\utilityStar[1]}{u^*_{#1}}
\newcommand{\throughput[1]}{T_{#1}}
\newcommand{\rate[1]}{R_{#1}}
\safemath{\infobits}{D}
\safemath{\totalbits}{M}
\safemath{\userno}{K}
\newcommand{\power[1]}{p_{#1}}
\newcommand{\powerStar[1]}{p^*_{#1}}
\newcommand{\SINR[1]}{\gamma_{#1}}
\newcommand{\SINRStar[1]}{\gamma^*_{#1}}
\safemath{\SINRStarInf}{\overline{\SINR[]}^*}
\newcommand{\efficiencyFunction[1]}{f(#1)}
\newcommand{\efficiencyFunctionPrime[1]}{f^\prime(#1)}

\safemath{\frameno}{N_f}
\safemath{\pulseno}{N_c}
\safemath{\gain}{N}
\newcommand{\channelresponse[2]}{c_{#1}(#2)}
\safemath{\chiptime}{T_c}
\safemath{\pathno}{L}
\safemath{\srake}{\pathno_S}
\safemath{\prake}{\pathno_P}
\newcommand{\delay[1]}{\tau_{#1}}
\newcommand{\pathgain[2]}{\alpha_{#1}^{(#2)}}
\newcommand{\rakecoeff[2]}{\beta_{#1}^{(#2)}}
\safemath{\processingMatrix}{\bG}

\safemath{\SP}{\text{SP}}
\safemath{\SI}{\text{SI}}
\safemath{\MAI}{\text{MAI}}
\newcommand{\channelgain[1]}{h_{#1}}
\newcommand{\hSP[1]}{h_{#1}^{(\SP)}}
\newcommand{\hSI[1]}{h_{#1}^{(\SI)}}
\newcommand{\hMAI[1]}{h_{#1}^{(\MAI)}}
\safemath{\varnoise}{\sigma^2}
\newcommand{\vectornorm}[1]{\left|\left|{#1}\right|\right|}
\newcommand{\vecpathgain[1]}{\bmalpha_{#1}}
\newcommand{\vecrakecoeff[1]}{\bmbeta_{#1}}
\newcommand{\matpathgain[1]}{\matA_{#1}}
\newcommand{\matrakecoeff[1]}{\matB_{#1}}
\newcommand{\matCoeffHsi}{\matPhi}
\newcommand{\coeffHsi[1]}{\phi_{#1}}

\safemath{\game}{G}
\safemath{\userset}{\setK}
\newcommand{\pmin[1]}{\underline{p}_{#1}}
\newcommand{\pmax[1]}{\overline{p}_{#1}}
\newcommand{\powerset[1]}{P_{#1}}
\newcommand{\powervect[1]}{\vecp_{#1}}
\newcommand{\SIratio[1]}{\gamma_{0,{#1}}}
\newcommand{\MAIratio[1]}{\zeta_{#1}}
\newcommand{\bestresponse[1]}{r_{#1}}
\newcommand{\bestresponsevect[1]}{\vecr(#1)}

\newcommand{\functionGamma[1]}{\Gamma\left({#1}\right)}
\safemath{\powerTimesHsp}{q}
\safemath{\varq}{\sigma^2_\powerTimesHsp}
\safemath{\meanq}{\eta_\powerTimesHsp}

\newcommand{\optimumcoeff[1]}{\xi_{#1}}

\newcommand{\MatpathProfile[1]}{\bD^{\alpha}_{#1}}
\newcommand{\MatrakeProfile[1]}{\bD^{\beta}_{#1}}
\newcommand{\MatpathC[1]}{\bC^{\alpha}_{#1}}
\newcommand{\MatrakeC[1]}{\bC^{\beta}_{#1}}
\safemath{\loadFactor}{\rho}
\newcommand{\functionPhi[1]}{\varphi\left({#1}\right)}
\newcommand{\functionNu[1]}{\nu\left({#1}\right)}
\newcommand{\vecW[1]}{\bmw^{(#1)}}
\newcommand{\coeffW[2]}{\bmw^{(#1)}_{#2}}
\safemath{\as}{\stackrel{a.s.}{\rightarrow}}
\newcommand{\functionTheta[2]}{\theta_{#1}\left({#2}\right)}
\newcommand{\functionThetaQuad[2]}{\theta^2_{#1}\left({#2}\right)}

\safemath{\Po}{P_o}

\begin{document}

\title{Energy-Efficient Power Control in Impulse Radio UWB Wireless Networks}

\author{Giacomo~Bacci\thanks
  {This work was supported in part by the U.S. Defense Advanced Research 
    Projects Agency under Grant No.~HR0011-06-1-0052.}, 
  Marco~Luise\thanks
  {G. Bacci and M. Luise are with the Dipartimento Ingegneria 
    dell'Informazione, Universit{\`a} di Pisa - 56122 Pisa, Italy
    (email: giacomo.bacci@iet.unipi.it, marco.luise@iet.unipi.it).},
  H.~Vincent~Poor\thanks
  {H. V. Poor is with the Department of Electrical Engineering, 
    Princeton University - 08544 Princeton, NJ, USA
    (email: poor@princeton.edu).}
  and~Antonia~M.~Tulino\thanks
  {A. M. Tulino is with the Dipartimento Ingegneria Elettronica e 
    Telecomunicazioni, Universit{\`a} degli Studi di Napoli Federico II -
    80125 Napoli, Italy (email: atulino@princeton.edu).} 
}

\maketitle

\begin{abstract}
In this paper, a game-theoretic model for studying power control for wireless
data networks in frequency-selective multipath environments is analyzed. The
uplink of an impulse-radio ultrawideband system is considered. The effects
of self-interference and multiple-access interference on the performance
of generic Rake receivers are investigated for synchronous systems. Focusing
on energy efficiency, a noncooperative game is proposed in which users in the
network are allowed to choose their transmit powers to maximize their own
utilities, and the Nash equilibrium for the proposed game is derived. It is
shown that, due to the frequency selective multipath, the noncooperative
solution is achieved at different signal-to-interference-plus-noise ratios,
depending on the channel realization and the type of Rake receiver
employed. A large-system analysis is performed to derive explicit
expressions for the achieved utilities. The Pareto-optimal (cooperative)
solution is also discussed and compared with the noncooperative approach.
\end{abstract}


\section{Introduction}\label{sec:intro}

\PARstart{A}{s the demand} for wireless services increases, the need for
efficient resource allocation and interference mitigation in wireless data
networks becomes more and more crucial. A fundamental goal of radio resource
management is transmitter power control, which aims to allow each user
to achieve the required quality of service (QoS) at the uplink receiver
without causing unnecessary interference to other users in the system.
Another key issue in wireless system design is energy consumption at user
terminals, since, in many scenarios, the terminals are battery-powered.

Recently, game theory has been used as an effective tool to study 
noncooperative power control in data networks \cite{mackenzie, goodman1, 
goodman2, saraydar1, saraydar2, feng, meshkati1, meshkati2}. The advantages 
of noncooperative (distributed) approaches with respect to a cooperative 
(centralized) approach are mainly due to the scalability of the network. Many 
of the problems to be solved in a communications system are in fact known to be
NP-hard. As a consequence, real-time solution of these optimization problems 
in a centralized fashion becomes infeasible as the network size increases and 
as the number of users varies \cite{mackenzie}. Game theory is the natural
framework for modeling and studying these kinds of interactions. A model for
power control as a noncooperative game is proposed in \cite{goodman1}. In this
scenario, the users choose their transmit powers to maximize their utilities, 
defined as the ratio of throughput to transmit power. In \cite{goodman2}, a 
network-assisted power-control scheme is proposed to improve the overall 
utility of a direct-sequence code-division multiple access (DS-CDMA) 
system. In \cite{saraydar1, saraydar2}, the authors use pricing to obtain
a more efficient solution for the power control game. 
Joint network-centric and user-centric power control are discussed in
\cite{feng}. In \cite{meshkati1}, the authors propose a power control
game for multicarrier CDMA (MC-CDMA) systems, while in \cite{meshkati2}
the effects of the receiver have been considered, particularly extending the
study to multiuser detectors and multiantenna receivers.

This work considers power control in ultrawideband (UWB) systems.
UWB technology is considered to be a potential candidate for
next-generation short-range high-speed data transmission, due to
its large spreading factor (which implies large multiuser capacity)
and low power spectral density (which allows coexistence with
incumbent systems in the same frequency bands). Commonly,
impulse-radio (IR) systems, which transmit very short pulses with a
low duty cycle, are employed to implement UWB systems \cite{win1,
win2}. In an IR system, a train of pulses is sent and the
information is usually conveyed by the position or the polarity of
the pulses, which correspond to pulse position modulation (PPM) and
binary phase shift keying (BPSK), respectively. To prevent
catastrophic collisions among different users and, thus, to provide
robustness against multiple access interference (MAI), each
information symbol is represented by a sequence of pulses; the
positions of the pulses within that sequence are determined by a
pseudo-random time-hopping (TH) sequence that is specific to each
user \cite{win2}. In ``classical'' impulse radio, the polarity of
those pulses representing an information symbol is always the same,
whether PPM or BPSK is employed \cite{win2, giannakis}. Recently,
pulse-based polarity randomization was proposed \cite{nakache},
where each pulse has a random polarity code in addition to the
modulation scheme, providing additional robustness against MAI
\cite{fishler1} and helping to optimize the spectral shape according
to US Federal Communications Commission (FCC) specifications
\cite{fcc}. Due to the large bandwidth, UWB signals have a much
higher temporal resolution than conventional narrowband or wideband
signals. Hence, channel fading cannot be assumed to be flat
\cite{molisch}, and self-interference (SI) must be taken into
account \cite{gezici2}. To the best of our knowledge, this paper is
the first to study the problem of radio resource allocation in a
frequency-selective multipath environment using a game-theoretic
approach. Previous work in this area has assumed flat fading
\cite{hayajneh, sun, huang}.

Our focus throughout this work is on energy efficiency. In this kind of
applications it is often more important to maximize the number of
bits transmitted per Joule of energy consumed than to maximize throughput.
We first propose a noncooperative (distributed) game in which users are
allowed to choose their transmit powers according to a utility-maximization
criterion. Focusing on Rake receivers \cite{proakis} at the base station,
we derive the Nash equilibrium for the proposed game, also proving its
existence and uniqueness. Using a large-system analysis, we obtain an 
approximation of the interference which is suitable for any kind of 
fading channels, including both small- and large-scale statistics. We also 
compute explicit expressions for the utility achieved at the equilibrium 
in a particular scenario and compare the performance of our
noncooperative approach with the optimal cooperative
(centralized) solution. It is shown that the difference between these
two approaches is not significant for typical networks.

The remainder of this paper is organized as follows. Some background
for this work is given in Sect. \ref{sec:problemStatement}. In Sect.
\ref{sec:npcg}, we describe our power control game and analyze the Nash
equilibrium for this game. In Sect. \ref{sec:ne}, we use the
game-theoretic framework along with a large-system analysis to evaluate
the performance of the system in terms of transmit powers and achieved
utilities. The Pareto-optimal (cooperative) solution to the power control
game is discussed in Sect. \ref{sec:pareto}, and its performance is compared
with that of the noncooperative approach. Numerical results are discussed in 
Sect. \ref{sec:simulation}, where also an iterative and distributed 
algorithm for reaching the Nash equilibrium is presented. Some conclusions 
are drawn in Sect. \ref{sec:conclusions}.

\section{Problem Statement}\label{sec:problemStatement}

\subsection{Formulation}\label{subsec:background}

Consider the uplink of an IR-UWB data network, where every user wishes to
locally and selfishly choose its action to maximize its
own utility function. The strategy chosen by a user affects the performance
of the other users in the network through MAI. Furthermore, since a realistic
IR-UWB transmission takes place in frequency-selective multipath channels,
the effect of SI cannot be neglected.

To pose the power control problem as a noncooperative game, a suitable 
definition of a utility function is needed to measure energy efficiency for
wireless data applications. A tradeoff relationship exists between obtaining 
high signal-to-interference-plus-noise ratio (SINR) levels and consuming 
low energy. These issues can be quantified \cite{goodman1} by defining the 
utility function of the $k$th user to be the ratio of its throughput
$\throughput[k]$ to its transmit power $\power[k]$, i.e.
\be
  \label{eq:utility}
  \utility[k]{\powervect[]}=\frac{\throughput[k]}{\power[k]},
\ee
where $\powervect[]=[\power[1],\dots,\power[\userno]]$ is the vector of
transmit powers, with \userno denoting the number of users in the network.
Throughput, here referred to as the net number of information bits that are 
received without error per unit time, can be expressed as
\be
  \label{eq:throughput}
  \throughput[k]=\frac{\infobits}{\totalbits}\rate[k] f_s(\SINR[k]),
\ee
where \infobits and \totalbits are the number of information bits and
the total number of bits in a packet, respectively; $\rate[k]$ and
$\SINR[k]$ are the transmission rate and the SINR for the $k$th user,
respectively; and $f_s(\SINR[k])$ is the efficiency function
representing the packet success rate (PSR), i.e., the probability that
a packet is received without an error. Our assumption is that a packet
will be retransmitted if it has one or more bit errors. The PSR depends
on the details of the data transmission, including its modulation, coding,
and packet size. To prevent the mathematical anomalies described in 
\cite{goodman1}, we replace PSR with an \emph{efficiency function}
$\efficiencyFunction[{\SINR[k]}]$ when calculating the throughput for 
our utility function. In the case of BPSK TH-IR systems in multipath 
fading channels, a reasonable approximation to the PSR for moderate-to-large
values of \totalbits is
$\efficiencyFunction[{\SINR[k]}]=(1-\text{e}^{-\SINR[k]/2})^\totalbits$. 

However, our analysis throughout this paper is valid for any efficiency
function that is increasing, S-shaped,\footnote{An increasing function 
is S-shaped if there is a point above which the function is concave, 
and below which the function is convex.} and continuously differentiable,
with $\efficiencyFunction[0]=0$, $\efficiencyFunction[+\infty]=1$,
and $\efficiencyFunctionPrime[0]=d\efficiencyFunction[{\SINR[k]}]/
d\SINR[k]|_{\SINR[k]=0}=0$. These assumptions are valid in many
practical systems. Furthermore, we assume that all users have the same
efficiency function. Generalization to the case where the efficiency
function is dependent on $k$ is straightforward. Note that the
throughput $\throughput[k]$ in (\ref{eq:throughput}) could also be
replaced with the Shannon capacity formula if $\utility[k]{\powervect[]}$
in (\ref{eq:utility}) is appropriately modified to ensure
$\utility[k]{\powervect[]}=0$ when $\power[k]=0$.

\begin{figure}
  \centering
  \includegraphics[width=12.0cm]{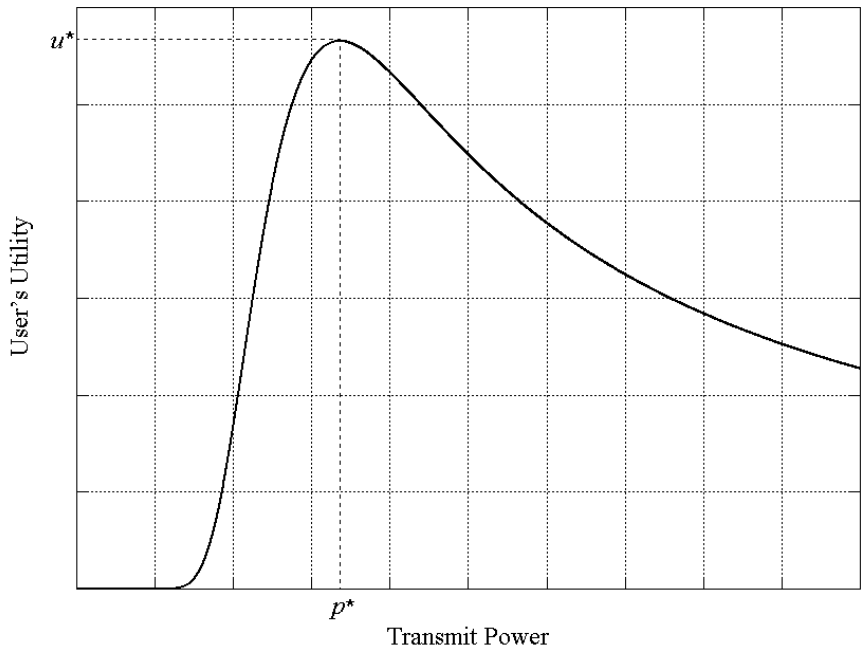}
  \caption{User's utility as a function of transmit power for fixed
    interference.}
  \label{fig:utility_shape}
\end{figure}

Combining (\ref{eq:utility}) and (\ref{eq:throughput}), and replacing the 
PSR with $\efficiencyFunction[{\SINR[k]}]$, 
\be
  \label{eq:utility2}
  \utility[k]{\powervect[]}=
  \frac{\infobits}{\totalbits}\rate[k]
  \frac{\efficiencyFunction[{\SINR[k]}]}{\power[k]}.
\ee

This utility function, which has units of bits/Joule, represents the
total number of data bits that are delivered to the destination without
an error per Joule of energy consumed, capturing the tradeoff between
throughput and battery life. For the sake of simplicity,
we assume that the transmission rate is the same for all users,
i.e., $\rate[1]=\dots=\rate[\userno]=\rate[]$. All the results obtained
here can easily be generalized to the case of unequal rates. Fig.
\ref{fig:utility_shape} shows the shape of the utility function in
(\ref{eq:utility2}) as a function of transmit power keeping other
users' transmit power fixed (the meaning of $\powerStar[]$ and
$\utilityStar[]$ will be provided in the remainder of the paper).

\subsection{System Model}\label{subsec:model}

We consider a BPSK random TH-IR system\footnote{Throughout all the paper,
we consider IR-UWB systems with polarity code randomization
\cite{nakache}.} with \userno users in the network transmitting
to a receiver at a common concentration point. The processing gain of
the system is assumed to be $\gain=\frameno\cdot\pulseno$, where \frameno
is the number of pulses that represent one information symbol, and \pulseno
denotes the number of possible pulse positions in a frame \cite{win1}.
The transmission is assumed to be over \emph{frequency selective channels},
with the channel for user $k$ modeled as a tapped delay line:
\be
  \label{eq:channel}
  \channelresponse[k]{t} =
  \sum_{l=1}^{\pathno}{\pathgain[l]{k}\delta(t-(l-1)\chiptime-\delay[k])},
\ee
where \chiptime is the duration of the transmitted UWB pulse, which is
the minimum resolvable path interval; \pathno is the number
of channel paths;
$\vecpathgain[k]=\tp{[\pathgain[1]{k},\dots,\pathgain[\pathno]{k}]}$ and
$\delay[k]$ are the fading coefficients and the delay of user $k$,
respectively. Considering a chip-synchronous scenario, the symbols are
misaligned by an integer multiple of the
chip interval \chiptime: $\delay[k] = \Delta_k\chiptime$, for
every $k$, where $\Delta_k$ is uniformly distributed in
$\{0,1,\dots,\gain-1\}$. In addition we assume that the channel
characteristics remain unchanged over a number of symbol 
intervals \cite{gezici2}.

Especially in indoor environments, multipath channels can have
hundreds of multipath components due to the high resolution of UWB signals.
In such cases, linear receivers such as matched filters (MFs),
pulse-discarding receivers \cite{fishler2}, and multiuser detectors (MUDs)
\cite{verdu} cannot provide good performance, since more collisions
will occur through multipath
components. In order to mitigate the effect of multipath fading as much
as possible, we consider a base station where \userno Rake receivers
\cite{proakis} are used.\footnote{For ease of calculation, perfect
channel estimation is considered throughout the paper.}
The Rake receiver for user $k$ is composed of \pathno fingers, where the
vector $\vecrakecoeff[k]=\processingMatrix\cdot\vecpathgain[k]=
\tp{[\rakecoeff[1]{k},\dots,\rakecoeff[\pathno]{k}]}$ represents the combining 
weights for user $k$, and the $\pathno\times\pathno$ matrix 
$\processingMatrix$ depends on the type of Rake receiver employed. 
In particular, if $\processingMatrix=\matIdentity$, an all-Rake 
(ARake) receiver is considered. 

The SINR of the $k$th user at the output of
the Rake receiver can be well approximated\footnote{This approximation
is valid for large \frameno (typically, at least 5).} by \cite{gezici2}
\be
  \label{eq:sinr}
  \SINR[k] = \frac{\hSP[k]\power[k]}{\displaystyle{\hSI[k]\power[k] +
      \sum_{\substack{j=1\\j\neq k}}^{\userno}{\hMAI[kj]\power[j]} +
      \sigma^2}},
\ee
where \varnoise is the variance of the additive white Gaussian
noise (AWGN) at the receiver, and the gains are expressed by
\begin{align}
  \label{eq:hSP}
  \hSP[k] &= \herm{\vecrakecoeff[k]}\cdot\vecpathgain[k],\\
  \label{eq:hSI}
  \hSI[k] &= \frac{1}{\gain}
  \frac{\vectornorm{\matCoeffHsi\cdot
      \left(\herm{\matrakecoeff[k]}\cdot\vecpathgain[k]+
      \herm{\matpathgain[k]}\cdot\vecrakecoeff[k]\right)}^2}
       {\herm{\vecrakecoeff[k]}\cdot\vecpathgain[k]},\\
  \label{eq:hMAI}
  \hMAI[kj] &= \frac{1}{\gain}
  \frac{\vectornorm{\herm{\matrakecoeff[k]}\cdot\vecpathgain[j]}^2
  + \vectornorm{\herm{\matpathgain[j]}\cdot\vecrakecoeff[k]}^2
  + \left|\herm{\vecrakecoeff[k]}\cdot\vecpathgain[j]\right|^2}
  {\herm{\vecrakecoeff[k]}\cdot\vecpathgain[k]},
\end{align}
where the matrices
\begin{align}
  \label{eq:matrixA}
  \matpathgain[k] &=
  \begin{pmatrix}
    \pathgain[\pathno]{k}&\cdots&\cdots&\pathgain[2]{k}\\
    0&\pathgain[\pathno]{k}&\cdots&\pathgain[3]{k}\\
    \vdots&\ddots&\ddots&\vdots\\
    0&\cdots&0&\pathgain[\pathno]{k}\\
    0&\cdots&\cdots&0
  \end{pmatrix},\\
  \label{eq:matrixB}
  \matrakecoeff[k] &=
  \begin{pmatrix}
    \rakecoeff[\pathno]{k}&\cdots&\cdots&\rakecoeff[2]{k}\\
    0&\rakecoeff[\pathno]{k}&\cdots&\rakecoeff[3]{k}\\
    \vdots&\ddots&\ddots&\vdots\\
    0&\cdots&0&\rakecoeff[\pathno]{k}\\
    0&\cdots&\cdots&0
  \end{pmatrix},\\
  \label{eq:matrixPhi}
  \matCoeffHsi &=
  \diag\left\{\coeffHsi[1],\dots,\coeffHsi[\pathno-1]\right\},
  \quad \coeffHsi[l]=\sqrt{\tfrac{\min\{\pathno-l,\pulseno\}}{\pulseno}},
\end{align}
have been introduced for convenience of notation.

By considering frequency selective channels, the transmit power of
the $k$th user, $\power[k]$, does appear not only in the numerator of
(\ref{eq:sinr}), but also in the denominator, owing to the SI due to
multiple paths. In the following sections, we extend the
approach of game theory to multipath channels, accounting for the SI in
addition to MAI and AWGN. The problem is more challenging
than with a single path since every user achieves a different
SINR at the output of its Rake receiver.

\section{The Noncooperative Power Control Game}\label{sec:npcg}

In this section, we propose a noncooperative power control game (NPCG)
in which every user seeks to maximize its own utility by choosing
its transmit power. Let
$\game = [\userset, \{\powerset[k]\}, \{\utility[k]{\powervect[]}\}]$
be the proposed noncooperative game where $\userset=\{1,\dots,\userno\}$
is the index set for the terminal users;
$\powerset[k]=\left[\pmin[k], \pmax[k]\right]$ is the strategy set,
with $\pmin[k]$ and $\pmax[k]$ denoting minimum and maximum power
constraints, respectively; and $\utility[k]{\powervect[]}$ is the
payoff function for user $k$ \cite{saraydar2}. Throughout this paper,
we assume $\pmin[k]=0$ and $\pmax[k]=\pmax[]>0$ for all $k\in\userset$.

Formally, the NPCG can be expressed as
\be
  \label{eq:pc_gen}
  \max_{\power[k]\in\powerset[k]} \utility[k]{\powervect[]} =
  \max_{\power[k]\in\powerset[k]} \utility[k]{\power[k], \powervect[-k]}
  \qquad \text{for $k=1,\dots,\userno$},
\ee
where $\powervect[-k]$ denotes the vector of transmit powers of all
terminals except terminal $k$.
The latter notation is used to emphasize that the $k$th user has
control over its own power $\power[k]$ only. Assuming equal transmission
rate for all users, (\ref{eq:pc_gen}) can be rewritten as
\be
  \label{eq:pc2}
  \max_{\power[k]\in\powerset[k]}
  \frac{\efficiencyFunction[{\SINR[k](\power[k], \powervect[-k])}]}
       {\power[k]} \qquad \text{for $k=1,\dots,\userno$},
\ee
where we have explicitly shown that $\SINR[k]$ is a function of
$\powervect[]$, as expressed in (\ref{eq:sinr}).

The solution that is most widely used for noncooperative game 
theoretic problems is the \emph{Nash equilibrium} \cite{mackenzie}. 
A Nash equilibrium is a set of strategies such that no user can 
unilaterally improve its own utility. Formally, a power vector
$\powervect[]=[\power[1],\dots,\power[\userno]]$ is a Nash equilibrium
of $\game = [\userset, \{\powerset[k]\}, \{\utility[k]{\powervect[]}\}]$
if, for every
$k\in\userset$, $\utility[k]{\power[k], \powervect[-k]}\ge
\utility[k]{\power[k]^\prime, \powervect[-k]}$ for all
$\power[k]^\prime\in\powerset[k]$.

The Nash equilibrium concept offers a predictable, stable outcome of
a game where multiple agents with conflicting interests compete
through self-optimization and reach a point where no player wishes to
deviate. However, such a point does not necessary exist. First, we
investigate the existence of an equilibrium in the NPCG.

\begin{theorem}\label{th:existence}
  A Nash equilibrium exists in the NPCG
  $\game = [\userset, \{\powerset[k]\}, \{\utility[k]{\powervect[]}\}]$.
  Furthermore, the unconstrained maximization of the utility function
  occurs when each user $k$ achieves an SINR $\SINRStar[k]$ that is 
  a solution of
  \be
    \label{eq:f_der}
    \efficiencyFunctionPrime[{\SINRStar[k]}]
    \SINRStar[k]\left(1-\SINRStar[k]/\SIratio[k]\right)=
    \efficiencyFunction[{\SINRStar[k]}],
  \ee
  where
  \be
    \label{eq:SIratio}
    \SIratio[k]=\frac{\hSP[k]}{\hSI[k]}
    =\gain\cdot
    \frac{(\herm{\vecrakecoeff[k]}\cdot\vecpathgain[k])^2}
      {\vectornorm{\matCoeffHsi\cdot
      \left(\herm{\matrakecoeff[k]}\cdot\vecpathgain[k]+
      \herm{\matpathgain[k]}\cdot\vecrakecoeff[k]\right)}^2}
    \ge1
  \ee
  and $\efficiencyFunctionPrime[{\SINRStar[k]}]=
  d\efficiencyFunction[{\SINR[k]}]/d\SINR[k]|_{\SINR[k]=\SINRStar[k]}$.
\end{theorem}
The proof of Theorem \ref{th:existence} can be found in 
App. \ref{pr:existence}.

The Nash equilibrium can be seen from another point of view. The power
level chosen by a \emph{rational} self-optimizing user constitutes a
\emph{best response} to the powers chosen by other players.
Formally, terminal $k$'s best response
$\bestresponse[k]: \powerset[-k]\rightarrow\powerset[k]$ is the
correspondence that assigns to each $\powervect[-k]\in\powerset[-k]$
the set
\begin{multline}
  \label{eq:bestresponse}
  \bestresponse[k](\powervect[-k])=\{\power[k]\in\powerset[k]:
  \utility[k]{\power[k], \powervect[-k]}\ge
  \utility[k]{\power[k]^\prime, \powervect[-k]}\}\\
  \text{for all $\power[k]^\prime\in\powerset[k]$},
\end{multline}
where $\powerset[-k]$ is the strategy space of all users excluding
user $k$.

With the notion of a terminal's best response correspondence, the
Nash equilibrium can be restated in a compact form: the
power vector $\powervect[]$ is a Nash equilibrium of the NPCG
$\game = [\userset, \{\powerset[k]\}, \{\utility[k]{\powervect[]}\}]$
if and only if $\power[k]\in\bestresponse[k](\powervect[-k])$ for
all $k\in\userset$.

\begin{proposition}
  \label{prop:bestResponse}
  Using the above definition in the NPCG, with a slight abuse of notation, 
  terminal $k$'s best response
  to a given interference vector $\powervect[-k]$ is
  \cite{saraydar2}
  \be
    \label{eq:prop1}
    \bestresponse[k](\powervect[-k])=\min(\pmax[],\powerStar[k]),
  \ee
  where
  \begin{align}
    \label{eq:power*}
    \powerStar[k]&=\argmax_{\power[k]\in\positivereals}{\utility[k]
    {\power[k],\powervect[-k]}}\nonumber\\
    &=\frac{\SINRStar[k]\left(\sum_{j\neq k}{\hMAI[kj]\power[j]}+
      \sigma^2\right)}{\hSP[k]\left(1-\SINRStar[k]/\SIratio[k]\right)}
  \end{align}
  is the unconstrained maximizer of the utility in (\ref{eq:utility2}) 
  (see Fig. \ref{fig:utility_shape}). Furthermore, $\powerStar[k]$ is unique.
  The proof of Prop. \ref{prop:bestResponse} can be found in App. 
  \ref{pr:bestResponse}.
\end{proposition}

It is worth noting that, at any equilibrium of the NPCG, a terminal
either attains the utility maximizing SINR $\SINRStar[k]$ or it fails to
do so and transmits at maximum power $\pmax[]$.

\begin{theorem}\label{th:uniqueness}
  The NPCG has a unique Nash equilibrium.
\end{theorem}
The proof of Theorem \ref{th:uniqueness} can be found in App. 
\ref{pr:uniqueness}.

\section{Analysis of the Nash equilibrium}\label{sec:ne}

\begin{figure}
  \centering
  \includegraphics[width=12.0cm]{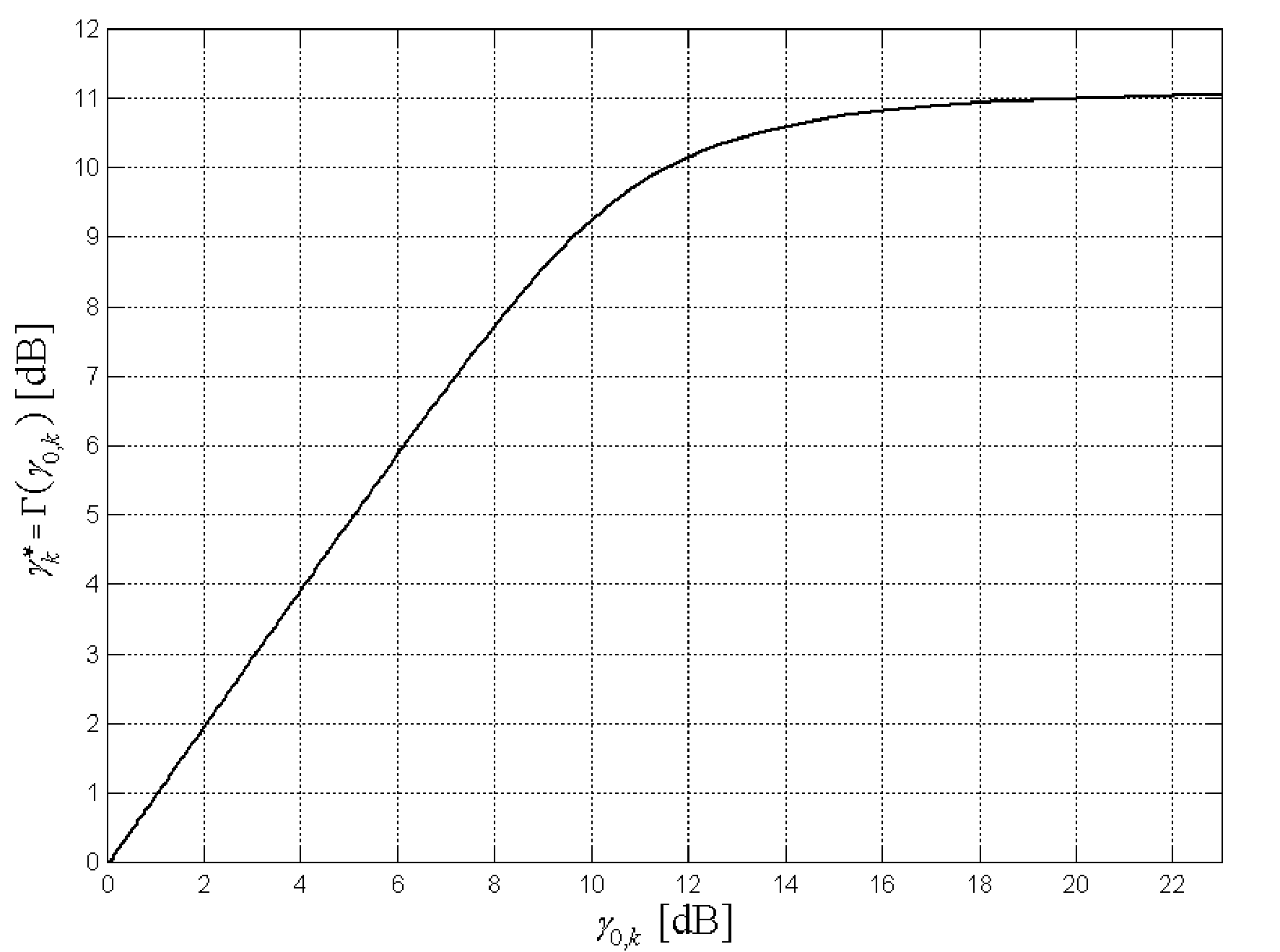}
  \caption{Shape of $\SINRStar[k]$ as a function of $\SIratio[k]$
    ($\totalbits=100$).}
  \label{fig:gammaStar}
\end{figure}

In the previous section, we have proven that a Nash equilibrium for
the NPCG exists and is unique. In the following, we study the properties
of this equilibrium. It is worth emphasizing that, unlike the previous
work in this area, $\SINRStar[k]$ is dependent on $k$, because of the SI
in (\ref{eq:sinr}). Hence, each user attains a different
level of SINR. More importantly, the only term dependent on $k$ in
(\ref{eq:f_der}) is $\SIratio[k]$, which is affected only by the channel
of user $k$. This means that $\SINRStar[k]$ can be assumed constant
when the channel characteristics remain unchanged, irrespectively of the
transmit powers $\powervect[]$ and the channel coefficients of the other
users. For convenience of notation, we can express $\SINRStar[k]$ as a
function of $\SIratio[k]$:
\be
  \label{eq:Gamma}
  \SINRStar[k]=\functionGamma[{\SIratio[k]}].
\ee
Fig. \ref{fig:gammaStar} shows the shape of $\SINRStar[k]$ as a function
of $\SIratio[k]$, where the efficiency function is taken as
$\efficiencyFunction[{\SINR[k]}]=(1-\text{e}^{-\SINR[k]/2})^\totalbits$,
with $\totalbits=100$. Even though $\SINRStar[k]$ is shown for values
of $\SIratio[k]$ approaching $0\,\text{dB}$, it is worth emphasizing that
$\SIratio[k]>10\,\text{dB}$ in most practical situations.

As can be noticed, the NPCG proposed herein represents a generalization
of the power control games discussed thoroughly in literature
\cite{goodman1, goodman2, saraydar1, saraydar2, feng, meshkati1, meshkati2}. 
If $\pathno=1$, i.e. in a flat-fading scenario, we obtain from 
(\ref{eq:hSI}) and (\ref{eq:SIratio}) that $\SIratio[k]=\infty$ for 
all $k$. This implies that $\SINRStar[k]$ is the same for every 
$k\in\userset$, and thus it is possible to apply the approach proposed, 
e.g., in \cite{saraydar2}.

\begin{assumption}\label{ass:q}
To simplify the analysis, let us assume the typical case of multiuser
UWB systems, where $\gain\gg\userno$. In addition, $\pmax[]$ is considered
sufficiently large that $\power[k]<\pmax[]$ for those users who
achieve $\SINRStar[k]$. In particular, when $\gain\gg\userno$, at the
Nash equilibrium the following property holds:
\be
  \label{eq:q}
  \hSP[k]\powerStar[k]\simeq\powerTimesHsp>0\qquad \forall k \in\userset.
\ee
The heuristic derivation of (\ref{eq:q}) can be justified by SI
reduction due to the hypothesis $\gain\gg\userno>1$.  Using
(\ref{eq:hSI}), $\SIratio[k]\gg 1$ for all $k$. Hence,
the noncooperative solution will be similar to that studied, e.g., in
\cite{meshkati2}. The validity of this assumption will be shown in
Sect. \ref{sec:simulation} through simulations.
\end{assumption}

The following proposition helps identify the Nash equilibrium for a
given set of channel realizations.

\begin{proposition}
  \label{prop:requirement}
  A necessary and sufficient condition for a desired SINR
  $\SINRStar[k]$ to be achievable is
  \be
    \label{eq:requirement}
    \SINRStar[k]\cdot\left(\SIratio[k]^{-1}+\MAIratio[k]^{-1}\right)<1
    \quad \forall k\in\userset,
  \ee
  where $\SIratio[k]$ is defined as in (\ref{eq:SIratio}), and
  \be
    \label{eq:MAIratio}
    \MAIratio[k]^{-1}=
    \sum_{\substack{j=1\\j\neq k}}^{\userno}{\frac{\hMAI[kj]}{\hSP[j]}}.
  \ee
  When (\ref{eq:requirement}) holds, each user can reach the optimum
  SINR, and the minimum power solution to do so is to assign each user
  $k$ a transmit power
  \be
    \label{eq:minimumPower}
    \powerStar[k]=\frac{1}{\hSP[k]}\cdot
    \frac{\sigma^2\SINRStar[k]}
     {1-\SINRStar[k]\cdot
       \left(\SIratio[k]^{-1}+\MAIratio[k]^{-1}\right)}.
  \ee
  When (\ref{eq:requirement}) does not hold, the users cannot achieve
  $\SINRStar[k]$ simultaneously, and some of them would end up
  transmitting at the maximum power $\pmax[]$. The proof of Prop. 
  \ref{prop:requirement} can be found in App. \ref{pr:requirement}.
\end{proposition}

Based on Prop. \ref{prop:requirement}, the amount of transmit power
$\powerStar[k]$ required to achieve the target SINR $\SINRStar[k]$ will
depend not only on the gain $\hSP[k]$, but also on the SI term $\hSI[k]$
(through $\SIratio[k]$) and the interferers $\hMAI[kj]$ (through
$\MAIratio[k]$). In order to derive some quantitative results for the
utility function and for the transmit powers independent of SI and
MAI terms, it is possible to resort to a large systems analysis. 
For convenience of notation, we introduce the following definitions, 
with $\Varop[\cdot]$ denoting the variance of a random variable:
\begin{itemize}
  \item let $\MatpathProfile[j]$ be a diagonal matrix whose elements are 
    \be\label{eq:MatpathProfile}
      \{\MatpathProfile[j]\}_{l} = \sqrt{\Varop[\pathgain[j]{l}]};
    \ee
  \item let $\MatrakeProfile[k]$ be a diagonal matrix whose elements are 
    \be\label{eq:MatrakeProfile}
      \{\MatrakeProfile[k]\}_{l} = \sqrt{\Varop[\rakecoeff[k]{l}]};
    \ee
  \item let $\MatpathC[j]$ be an $\pathno\times(\pathno-1)$ matrix whose 
    elements are 
    \be\label{eq:MatpathC}
      \{\MatpathC[j]\}_{li}=
      \sqrt{\frac{\Varop[\{\matpathgain[j]\}_{li}]}{\pathno}};
    \ee
  \item let $\MatrakeC[j]$ be an $\pathno\times(\pathno-1)$ matrix whose 
    elements are 
    \be\label{eq:MatrakeC}
      \{\MatrakeC[k]\}_{li}=
      \sqrt{\frac{\Varop[\{\matrakecoeff[k]\}_{li}]}{\pathno}};
    \ee
  \item let $\functionPhi[\cdot]$ be the matrix operator
    \be
      \label{eq:Phi}
      \functionPhi[\cdot]=
      \lim_{\pathno\rightarrow\infty}{\frac{1}{\pathno}\Tr(\cdot)}, 
    \ee
    where $\Tr(\cdot)$ is the trace operator.
\end{itemize}

\begin{theorem}\label{th:zeta}
  Assume that $\pathgain[k]{l}$ are zero-mean random variables independent 
  across $k$ and $l$, and $\processingMatrix$ is a deterministic diagonal 
  matrix (thus implying that $\pathgain[k]{l}$ and $\rakecoeff[j]{m}$ are 
  dependent only when $j=k$ and $m=l$). In the asymptotic case where \userno 
  and \frameno are finite,\footnote{In order for the analysis to be 
    consistent, and also considering regulations by the FCC \cite{fcc}, it 
    is worth noting that \frameno could not be smaller than a certain 
    threshold ($\frameno\ge5$).} while $\pathno, \pulseno \rightarrow \infty$, 
  the term $\MAIratio[k]^{-1}$ converges almost surely (a.s.) to 
  \be\label{eq:zetaTh}
    \MAIratio[k]^{-1}\as
    \frac{1}{\gain}\sum_{\substack{j=1\\j\neq k}}^{\userno}
         {\frac{\functionPhi[{\MatpathProfile[j]\MatrakeC[k]
                 \herm{{\MatrakeC[k]}}\MatpathProfile[j]}] +
             \functionPhi[{\MatrakeProfile[k]\MatpathC[j]
                 \herm{{\MatpathC[j]}}\MatrakeProfile[k]}]}
           {\functionPhi[{\MatpathProfile[j]\MatrakeProfile[j]}]\cdot
             \functionPhi[{\MatpathProfile[k]\MatrakeProfile[k]}]}}.
  \ee
\end{theorem}

\begin{theorem}\label{th:gamma}
  Assume $\pathgain[k]{l}$ and $\processingMatrix$ as in Theorem 
  \ref{th:zeta}. In the asymptotic case where \userno and \frameno are 
  finite, while $\pathno, \pulseno \rightarrow \infty$, 
  \be\label{eq:gammaTh}
    \SIratio[k]^{-1} \as \frac{1}{\gain}
    \frac{\displaystyle{\lim_{\pathno\rightarrow\infty}{\frac{1}{\pathno^2}
          \sum_{i=1}^{\pathno-1}{\coeffHsi[i]^2}\cdot
          \sum_{l=1}^{i}{\functionThetaQuad[k]{l,\pathno+l-i}}}}}
         {\left(\functionPhi[{\MatpathProfile[k]\MatrakeProfile[k]}]\right)^2},
  \ee
  where $\coeffHsi[i]$ is defined as in (\ref{eq:matrixPhi}) and
  \begin{align}\label{eq:Theta}
    \functionTheta[k]{l,\pathno+l-i}&=
    \{\MatpathProfile[k]\}_l \{\MatrakeProfile[k]\}_{\pathno+l-i}\nonumber\\
    &+ \{\MatrakeProfile[k]\}_l \{\MatpathProfile[k]\}_{\pathno+l-i}.
  \end{align}
\end{theorem}
The proofs of Theorems \ref{th:zeta} and \ref{th:gamma} can be found in App. 
\ref{sec:proofs_3_4}.

It is worth emphasize that the results above can be applied to any kind of 
fading models, since only the second-order statistics are required. 
Furthermore, due to the symmetry of (\ref{eq:zetaTh}) and (\ref{eq:gammaTh}), 
it is easy to verify that the results are independent of large-scale fading 
models. Hence, Theorems \ref{th:zeta} and \ref{th:gamma} apply to any kind 
of channel, which may include both large- and small-scale statistics.

For ease of calculation, in the following we derive the asymptotic values 
when considering a flat Power Delay Profile (PDP) \cite{rappaport} for the 
channel coefficients. In addition, the variance of $\pathgain[k]{l}$ is 
assumed dependent on the user $k$, but independent of $l$, i.e., 
$\Varop[\pathgain[k]{l}]=\sigma^2_k$ for all $l$.

\begin{proposition}\label{prop:aRake}
  Under the above mentioned hypotheses, when using an ARake, and thus 
  $\processingMatrix=\matIdentity$, 
  \begin{align}\label{eq:zetaaRake}
    \MAIratio[k]^{-1}&\as\frac{\userno-1}{\gain},\\
    \label{eq:gammaaRake}
    \SIratio[k]^{-1}&\as\frac{\functionNu[\loadFactor]}{\gain},
  \end{align}
  where $\loadFactor=\pulseno/\pathno$, $0<\loadFactor<\infty$, and 
  \be
    \label{eq:functioNu}
    \functionNu[\loadFactor]=\begin{cases}
    \dfrac{2}{3}\left(3-3\loadFactor+\loadFactor^2\right),&\loadFactor\le1,\\
    2/(3\loadFactor),&\loadFactor>1.\end{cases}
  \ee
\end{proposition}
Using definitions (\ref{eq:matrixPhi}) and (\ref{eq:MatpathProfile}) -- 
(\ref{eq:Phi}), and applying Theorems \ref{th:zeta} and \ref{th:gamma}, 
after some algebraic manipulations, the proof is straightforward. As 
already noticed to justify Ass. \ref{ass:q}, but also from 
(\ref{eq:gammaaRake}), $\SIratio[k]\gg1$ for all $k$. Thus, 
$\SINRStar[k]=\functionGamma[{\SIratio[k]}]$ approaches 
$\SINRStarInf=\functionGamma[\infty]$, leading to a nearly 
SINR-balancing scenario. 

An immediate consequence of Prop. \ref{prop:aRake} is the expression for 
transmit powers and utility functions at the Nash equilibrium, which are 
independent of the channel realizations of the other users, and of the 
SI:\footnote{Of course, the amount of transmit power $\powerStar[k]$ 
  needed to achieve $\SINRStar[k]$ is dependent on the channel realization.}
\begin{align}
  \label{eq:power*LSA}
  \powerStar[k]&\simeq\frac{1}{\hSP[k]}\cdot\frac{\sigma^2\SINRStarInf}
  {1-\SINRStarInf\cdot[\userno-1+\functionNu[\loadFactor]]/\gain},\\
  \label{eq:utilityLSA}
  \utilityStar[k]&\simeq\hSP[k]\cdot\frac{\infobits}{\totalbits}\rate[k]
  \frac{\efficiencyFunction[{\SINRStarInf}]
    \left(1-\SINRStarInf\cdot[\userno-1+\functionNu[\loadFactor]]/\gain\right)}
       {\sigma^2\SINRStarInf},
\end{align}
where $\utilityStar[k]$ is the utility function of the user $k$ at
the Nash equilibrium (see Fig. \ref{fig:utility_shape}), and where the 
condition (\ref{eq:requirement}) translates into 
\be\label{eq:requirementLSA}
  \frameno\ge\left\lceil\SINRStarInf\cdot
  \frac{\userno-1+\functionNu[\loadFactor]}
       {\pulseno}\right\rceil.
\ee

The validity of the claims above is verified in Sect. \ref{sec:simulation} 
using simulations. To show that the results (\ref{eq:zetaaRake}) -- 
(\ref{eq:gammaaRake}) represent good approximations not only for the model 
with flat PDP and equal variances, which has been used only for convenience 
of calculation, simulations are carried out using the exponential decaying 
PDP described in \cite{cassioli}, which provides a more realistic channel 
model for the UWB scenario.

\section{Social Optimum}\label{sec:pareto}
The solution to the power control game is said to be Pareto-optimal if
there exists no other power allocation $\powervect[]$ for which one or
more users can improve their utilities without reducing the utility
of any of the other users. It can be shown that the Nash equilibrium
presented in the previous section is not Pareto-optimal. This means that
it is possible to improve the utility of one or more users without
harming other users. On the other hand, it can be shown that the solution
to the following social problem gives the Pareto-optimal frontier
\cite{meshkati2}:
\be
  \label{eq:optimum1}
  \powervect[opt]= \argmax_{\powervect[]} \sum_{k=1}^{\userno}
  {\optimumcoeff[k] \utility[k]{\powervect[]}},
\ee
for $\optimumcoeff[k]\in\positivereals$ (the set of positive real numbers). 
Pareto-optimal solutions are, in general, difficult to obtain. 
Here, we conjecture that the Pareto-optimal
solution occurs when all users achieve the same SINRs, $\SINR[opt]$.
This approach is chosen not only because SINR balancing ensures fairness
among users in terms of throughput and delay \cite{meshkati2}, but also
because, for large systems, the Nash equilibrium is achieved when all SINRs
are similar. We also consider the hypothesis
$\optimumcoeff[1]=\dots=\optimumcoeff[\userno]=1$, suitable for a
scenario without priority classes. Hence, the maximization
(\ref{eq:optimum1}) can be written as
\be
  \label{eq:optimum2}
  \powervect[opt]=
  \argmax_{\powervect[]} \efficiencyFunction[{\SINR[]}]\sum_{k=1}^{\userno}
  {\frac{1}{\power[k]}}.
\ee

In a network where the hypotheses of Ass. \ref{ass:q}, Theorem 
\ref{th:zeta} and Theorem \ref{th:gamma} are fulfilled, and where 
ARake receivers are employed, at the Nash equilibrium all users 
achieve a certain output SINR $\SINR[]$ with 
$\hSP[k]\power[k]\simeq\powerTimesHsp\left(\SINR[]\right)$, where
\be
  \label{eq:optimumq}
  \powerTimesHsp\left(\SINR[]\right)=
  \frac{\sigma^2\SINR[]}
       {1-\SINR[]\cdot[\userno-1+\functionNu[\loadFactor]]/\gain},
\ee
with $\loadFactor=\pulseno/\pathno$. Therefore, (\ref{eq:optimum2}) 
can be expressed as
\be
  \label{eq:optimum3}
  \SINR[opt]\simeq\argmax_{\SINR[]} \frac{\efficiencyFunction[{\SINR[]}]}
       {\powerTimesHsp\left(\SINR[]\right)}
  \sum_{k=1}^{\userno}{\hSP[k]},
\ee
since there exists a one-to-one correspondence between $\SINR[]$ and 
$\powervect[]$. It should be noted that, while the maximizations in 
(\ref{eq:pc2}) consider no cooperation among users, (\ref{eq:optimum2}) 
assumes that users cooperate in choosing their transmit powers. That means 
that the relationship between the user's SINR and transmit power will be 
different from that in the noncooperative case.

\begin{proposition}\label{prop:so}
  In a network where $\pathno, \pulseno\rightarrow\infty$ and 
  $\gain\gg\userno$, using ARake receivers, the Nash equilibrium 
  approaches the Pareto-optimal solution.
\end{proposition}

\begin{proof}\label{pr:so}
  The solution $\SINR[opt]$ to (\ref{eq:optimum3}) must satisfy the condition 
  $d\left(\efficiencyFunction[{\SINR[]}]/
  \powerTimesHsp\left(\SINR[]\right)\right)/
  d\SINR[]|_{\SINR[]=\SINR[opt]}=0$. Using this fact, combined with 
  (\ref{eq:optimumq}), gives us the equation that must be satisfied by the 
  solution of the maximization problem in (\ref{eq:optimum3}):
  \be
    \label{eq:optimumf}
    \efficiencyFunctionPrime[{\SINR[opt]}]\SINR[opt]\left[1-\SINR[opt]\cdot
      \frac{\userno-1+\functionNu[\loadFactor]}{\gain}\right]=
    \efficiencyFunction[{\SINR[opt]}].
  \ee 
  We see from (\ref{eq:optimumf}) that the Pareto-optimal solution differs 
  from the solution (\ref{eq:f_der}) of the noncooperative 
  utility-maximizing method, since (\ref{eq:optimumf}) also takes into 
  account the contribution of the interferers. In particular,
  \be
    \label{eq:Gamma2}
    \SINR[opt]=\functionGamma[{\frac{\userno-1+\functionNu[\loadFactor]}
        {\gain}}].
  \ee

  Since the function $\functionGamma[\cdot]$ is increasing with its 
  argument for any S-shaped $\efficiencyFunction[{\SINR[]}]$ (as can also 
  be seen in Fig. \ref{fig:gammaStar}), and since 
  $\gain/\left[(\userno-1+\functionNu[\loadFactor])\right]\le\SIratio[k]$ 
  for all $k$ (from (\ref{eq:zetaaRake}) and (\ref{eq:gammaaRake})),
  \be
    \label{eq:NEvsSO}
    \SINR[opt]\le\SINR[]\le\SINRStarInf, 
  \ee
  due to (\ref{eq:Gamma}) and (\ref{eq:Gamma2}). On the other hand, assuming 
  $\gain\gg\userno$ and $0<\loadFactor<\infty$ implies 
  $\SINR[opt]\rightarrow\SINRStarInf$. From (\ref{eq:NEvsSO}), it is 
  apparent that $\SINR[]\rightarrow\SINRStarInf$ as well. This means that, 
  in almost all typical scenarios, the target SINR for the noncooperative 
  game, $\SINR[]$, is close to the target SINR for the Pareto-optimal 
  solution, $\SINR[opt]$. Consequently, the average utility provided by 
  the Nash equilibrium is close to the one achieved according to the 
  Pareto-optimal solution.
\end{proof}

The validity of the above claims will be verified in Sect. 
\ref{sec:simulation} using simulations.

\section{Simulation Results}\label{sec:simulation}

\subsection{Implementation}\label{subsec:algorithm}
In this subsection, we present an iterative and distributed algorithm for
reaching a Nash equilibrium of the proposed power control game. This
algorithm is applicable to all types of Rake receivers, as well as to
any kind of channel model. The description of the algorithm is as
follows.
\\
\\
\emph{The Best-Response Power-Control (BRPC) Algorithm}
\\
\\
Consider a network with \userno users, a processing gain
$\gain=\frameno\cdot\pulseno$, a channel with \pathno fading paths,
and a maximum transmit power $\pmax[]$.
\begin{enumerate}
\item Simulate the channel fading coefficients $\vecpathgain[k]$ for
all users according to the chosen channel model.
\item Set the Rake receivers coefficients $\vecrakecoeff[k]$ for all
users according to the chosen receiver.
\item Compute the SP term, $\hSP[k]$, the SI term, $\hSI[k]$, the MAI
term, $\hMAI[kj]$, according to (\ref{eq:hSP}) -- (\ref{eq:hMAI}),
and the optimum SINR, $\SINRStar[k]$, solution of (\ref{eq:f_der}),
for all users.
\item Initialize randomly the transmit powers of all users within the
range $\left[0,\pmax[]\right]$.
\item Compute the received SINR $\SINR[k]$ at the base station for
  each user according to (\ref{eq:sinr}).
\item Set $k=1$.
\item Adjust the transmit power for user $k$ according 
  to (\ref{eq:prop1}) and to
  \be\label{eq:updatePower}
    \powerStar[k] = \power[k]\cdot
    \frac{\SINRStar[k]}{\SINR[k]}\cdot
    \frac{1-\SINR[k]/\SIratio[k]}{1-\SINRStar[k]/\SIratio[k]},
  \ee
  where $\SIratio[k]$ is defined as in (\ref{eq:SIratio}).
\item $k=k+1$.
\item If $k\le\userno$, then go back to Step 7.
\item Stop if the powers have converged; otherwise, go to Step 5.
\end{enumerate}

This is a best-response algorithm, since at each stage a user decides
to transmit at a power that maximizes its own utility (i.e., its
best-response strategy), given the current conditions of the system.
Looking at Step 7, it may appear 
from (\ref{eq:power*}) that the each user should know its own transmit 
power $\power[k]$ and ratio $\SIratio[k]$, as well as some other 
quantities ($\power[j]$ and $\hMAI[kj]$ for $j \neq k$) relevant to 
all of the other users in the network. On the contrary, it turns out that
user $k$ only needs to know its own received SINR at the 
base station $\SINR[k]$. In fact, the term due to interference-plus-noise 
in (\ref{eq:power*}) can be obtained from (\ref{eq:sinr}) as
\be\label{eq:estIN}
  \sum_{j\neq k}{\hMAI[kj]\power[j]}+ \sigma^2 = 
  \hSP[k]\power[k]\cdot\frac{1-\SINR[k]/\SIratio[k]}{\SINR[k]},
\ee
where $\power[k]$ is the transmit power of user $k$ at the previous 
iteration. Therefore, after straightforward manipulation, (\ref{eq:power*}) 
translates into the noncooperative update (\ref{eq:updatePower}).
The received SINR $\SINR[k]$ can be fed back to the user terminal from the 
access point, along with SP and SI terms.

It is clear that the above algorithm converges to a Nash equilibrium,
whose existence and uniqueness are proven in App. \ref{sec:proofs1_2}. 
The convergence of the BRPC algorithm is
also validated in Sect. \ref{subsec:results} using extensive simulations.

\subsection{Numerical Results}\label{subsec:results}

In this subsection, we present numerical results for the analysis presented
in the previous sections. We assume that each packet contains
$100\,\text{b}$ of
information and no overhead (i.e., $\infobits=\totalbits=100$). The
transmission rate is $\rate[]=100\,\text{kb/s}$, the thermal noise
power is $\sigma^2=5 \times 10^{-16}\,\text{W}$, and the maximum transmit
power is $\pmax[]=1\,\mu\text{W}$. We use the efficiency function
$\efficiencyFunction[{\SINR[k]}]=(1-\text{e}^{-\SINR[k]/2})^\totalbits$.
Using $\totalbits=100$, $\SINRStarInf=\functionGamma[\infty]=
11.1\,\text{dB}$. To model the UWB scenario \cite{molisch},
channel gains are simulated following \cite{cassioli}, where both
small- and large-scale statistics are taken into account. The distance 
between the users and the base station is assumed to be uniformly 
distributed between $3$ and $20\,\text{m}$.

\begin{table}
  \renewcommand{\arraystretch}{1.3}
  \caption{Ratio $\varq/\meanq^2$ for different network parameters.}
  \label{tab:q}
  \centering
  \begin{tabular}{c|c|c|c|c|}
    & \multicolumn{4}{|c|}{$(\pathno, \userno)$}\\
    \hline
    $(\pulseno, \frameno)$ & 
    \it{(20,8)} & \it{(20,16)} & \it{(50,8)} & \it{(50,16)} \\
    \hline
    \it{(30,10)} & 9.4E-4 & 3.2E-3 & 4.8E-4 & 1.7E-3 \\
    \hline
    \it{(30,50)} & 2.9E-5 & 6.4E-5 & 1.6E-5 & 3.4E-5 \\
    \hline
    \it{(50,10)} & 2.9E-4 & 6.8E-4 & 1.5E-4 & 3.7E-4 \\
    \hline
    \it{(50,50)} & 1.0E-5 & 2.2E-5 & 0.6E-5 & 1.2E-5 \\
    \hline
    \it{(100,10)} & 6.7E-5 & 1.5E-4 & 3.7E-5 & 7.8E-5 \\
    \hline
    \it{(100,50)} & 0.3E-5 & 0.6E-5 & 0.1E-5 & 0.3E-5 \\
    \hline
  \end{tabular}
\end{table}

Before showing the numerical results for both the noncooperative and the 
cooperative approaches, some simulations are provided to verify 
the validity of Ass. \ref{ass:q} introduced in Sect. \ref{sec:ne}. 
Table \ref{tab:q} reports the ratio $\varq/\meanq^2$ of the variance $\varq$ 
to the squared mean value $\meanq^2$ of the values 
$\powerTimesHsp=\hSP[k]\powerStar[k]$, obtained averaging $10\,000$ 
realizations of channel coefficients for different values of network 
parameters using ARake receivers. We can see that, when the processing 
gain is much greater than the number of users, $\varq/\meanq^2\ll1$. 
Hence, (\ref{eq:q}) can be used to carry out the theoretical analysis of the 
Nash equilibrium.

\begin{figure}
  \centering
  \includegraphics[width=12.0cm]{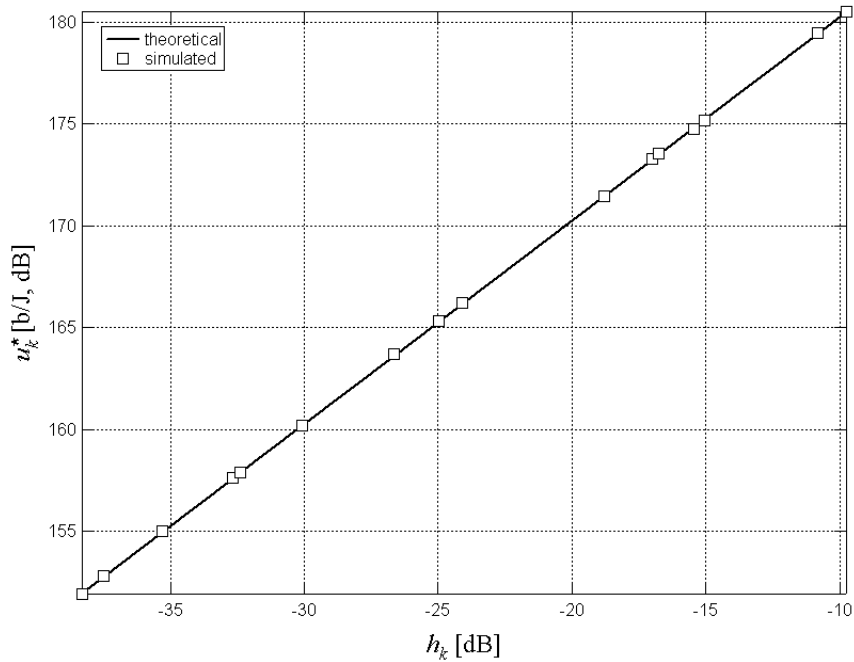}
  \caption{Achieved utility versus channel gain at the 
    Nash equilibrium for the ARake receiver.}
  \label{fig:utilityRake}
\end{figure}

Fig. \ref{fig:utilityRake} shows the utilities achieved at the Nash 
equilibrium as functions of the channel gains $\channelgain[k]=
\vectornorm{\vecpathgain[k]}^2$. These results have been 
obtained using random channel realizations for $\userno=16$ users. 
The number of possible pulse positions is $\pulseno=100$, while the 
number of paths is $\pathno=60$, in order to satisfy the large system 
assumption with $\functionNu[\loadFactor]=0.4$. The number of frame is 
$\frameno=10$, thus leading to a processing gain $\gain=1000\gg\userno$. 
The line represents the theoretical values of (\ref{eq:utilityLSA}) when 
using an ARake, whereas the square markers correspond to the simulations 
achieved with the BRPC algorithm. We can see that the simulations match 
closely with the theoretical results.

\begin{figure}
  \centering
  \includegraphics[width=12.0cm]{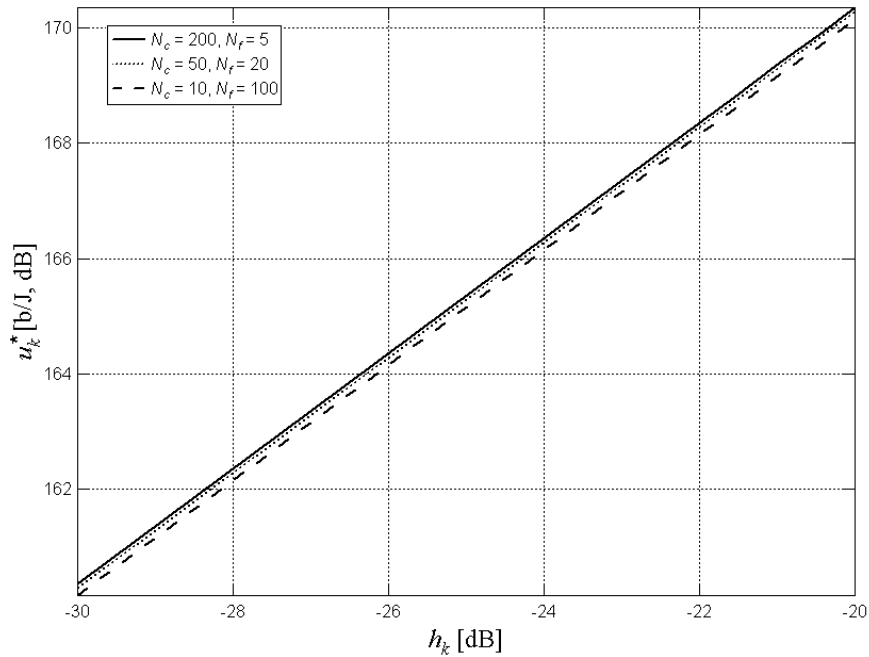}
  \caption{Achieved utility versus channel gain at the Nash equilibrium 
    for different ratios $\pulseno/\frameno$.}
  \label{fig:utilityNfixed}
\end{figure}

Fig. \ref{fig:utilityNfixed} shows the utility as a function of the 
channel gain when the processing gain \gain is constant, but the ratio 
$\pulseno/\frameno$ is variable.\footnote{To avoid a too busy graph, 
  only the theoretical values are reported.} The results have been 
obtained for a network with $\userno=16$ users, $\pathno=100$ channel paths 
and processing gain $\gain=1000$, using ARake receivers at the base station. 
The solid line corresponds to $\pulseno/\frameno=40$, the dotted line 
represents $\pulseno/\frameno=2.5$, and the dashed line shows 
$\pulseno/\frameno=0.1$. As expected, higher $\pulseno/\frameno$ ratios 
(and thus higher \pulseno, when \gain is fixed) correspond to higher 
utility, as described in (\ref{eq:utilityLSA}). In fact, 
$\functionNu[\loadFactor]$ decreases as $\loadFactor$ increases, i.e., as 
$\pulseno$ increases (when $\pathno$ is fixed). This result complies with 
theoretical analysis of UWB systems \cite{gezici2}, since, 
for a fixed total processing gain \gain, increasing the number of chips 
per frame, \pulseno, will decrease the effects of SI, while the 
dependency of the expressions on the MAI remains unchanged. Hence, a 
system with a higher \pulseno achieves better performance.

\begin{figure}
  \centering
  \includegraphics[width=12.0cm]{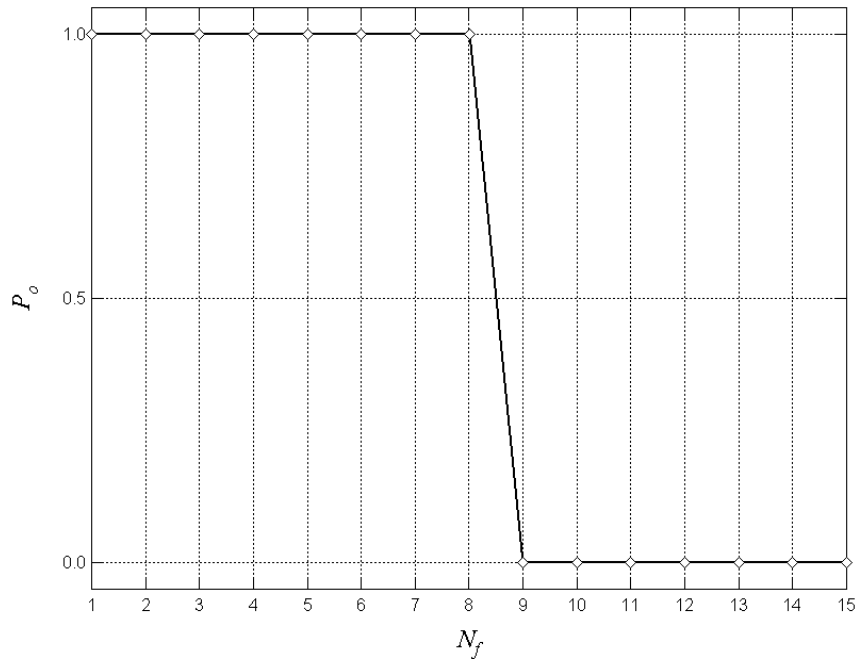}
  \caption{Probability of having at least one user transmitting at 
    maximum power versus number of frames.}
  \label{fig:Nfmin}
\end{figure}

Fig. \ref{fig:Nfmin} shows the probability 
$\Po=\Pr\{\max_k\power[k]=\pmax[]=1\,\mu\text{W}\}$ of having at least 
one user transmitting at the maximum power, as a function of 
the number of frames \frameno. We consider $10\,000$ realizations of the 
channel gains, using a network with ARake receivers at the base station, 
$\userno=32$ users, $\pulseno=50$, and $\pathno=100$ (thus 
$\loadFactor=0.5$ and $\functionNu[\loadFactor]\simeq 1. 17$). From 
(\ref{eq:requirementLSA}), the minimum value of 
\frameno that allows all \userno users to achieve the optimum SINRs is 
$\frameno=\lceil 8.33 \rceil = 9$. The simulations thus agree with the 
analytical results of Sect. \ref{sec:ne}.

\begin{figure}
  \centering
  \includegraphics[width=12.0cm]{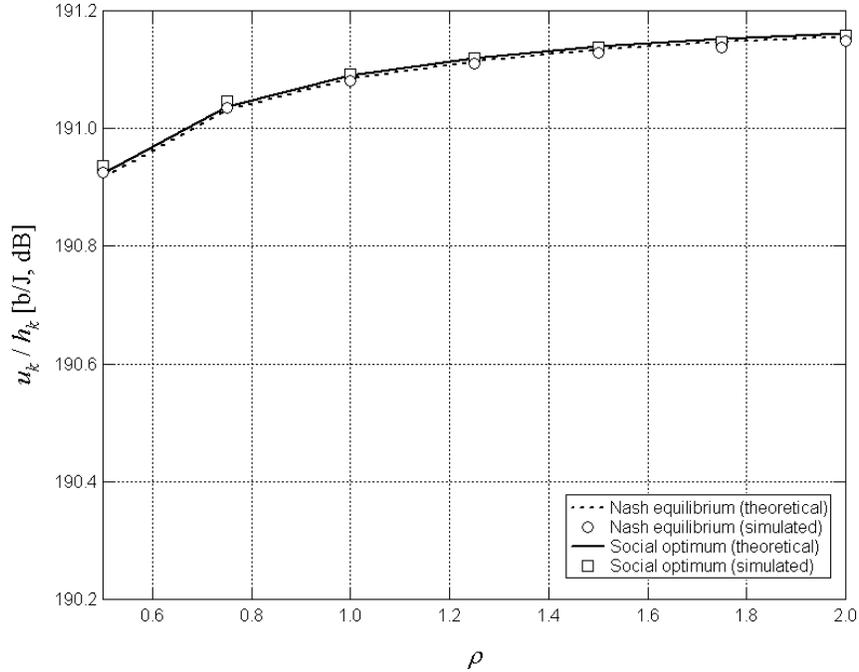}
  \caption{Comparison of the normalized utility versus 
    load factor for the noncooperative and Pareto-optimal solutions.}
  \label{fig:utilityNEvsSO}
\end{figure}

\begin{figure}
  \centering
  \includegraphics[width=12.0cm]{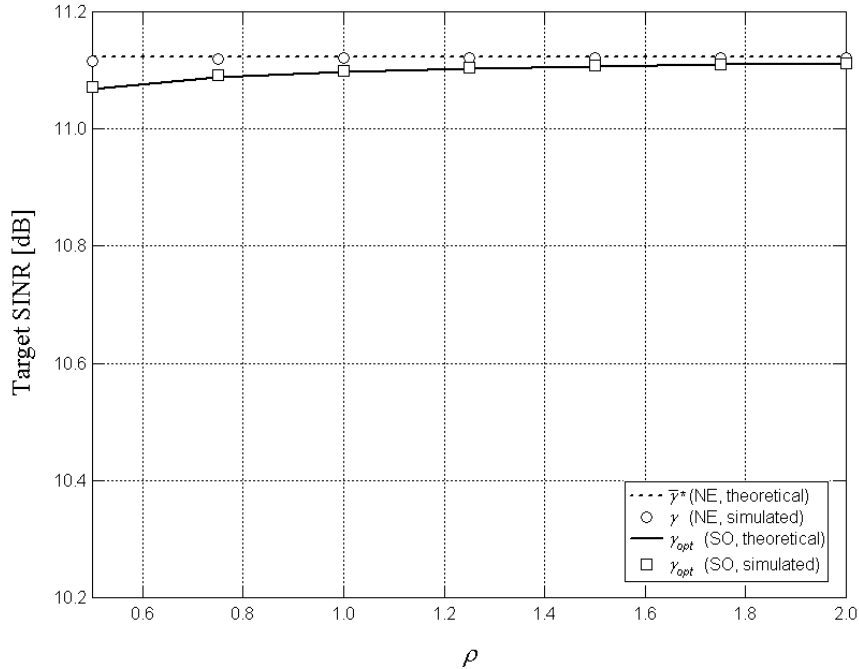}
  \caption{Comparison of the target SINRs versus 
    load factor for the noncooperative and Pareto-optimal solutions.}
  \label{fig:gammaNEvsSO}
\end{figure}

We now analyze the performance of the system when using a Pareto-optimal 
solution instead of the Nash equilibrium. Fig. \ref{fig:utilityNEvsSO} 
shows the normalized utility $u_k/\channelgain[k]$ as a function of the 
load factor $\loadFactor$. We consider a network with $\userno=5$ users, 
$\frameno=20$ frames and ARake receivers at the base station. The lines 
represent theoretical values of Nash equilibrium (dotted line), using 
(\ref{eq:utilityLSA}), and of the social optimum solution (solid line), 
using (\ref{eq:utilityLSA}) again, but substituting \SINRStarInf with the 
numerical solution of (\ref{eq:optimumf}), $\SINR[opt]$. The markers 
correspond to the simulation results. In particular, the circles represent 
the averaged solution of the BRPC algorithm, while the square markers show 
averaged numerical results (through a complete search) of the maximization 
(\ref{eq:optimum1}), with $\optimumcoeff[k]=1$. As stated in Sect. 
\ref{sec:pareto}, the difference between the noncooperative approach and 
the Pareto-optimal solution is not significant. Fig. \ref{fig:gammaNEvsSO} 
compares the target SINRs of the noncooperative solutions with the 
target SINRs of the Pareto-optimal solutions. As before, the lines 
correspond to the theoretical values (dashed line for the noncooperative 
solution, solid line for the social optimum solution), while the markers 
represent the simulation results (circles for the noncooperative solutions, 
square markers for the Pareto-optimal solution). It is seen that, in both 
cases, the averaged target SINRs for the Nash equilibrium, $\SINR[]$, and 
the averaged target SINRs for the social optimum solution, $\SINR[opt]$, 
are very close to \SINRStarInf, as shown in Sect. \ref{sec:pareto}.

\section{Conclusion}\label{sec:conclusions}

In this paper, we have used a game-theoretic approach to study 
power control for a wireless data network in frequency-selective environments, 
where the user terminals transmit IR-UWB signals and the common concentration 
point employs Rake receivers. A noncooperative game has been
proposed in which users are allowed to choose their transmit powers
according to a utility-maximizing criterion, where the utility function
has been defined as the ratio of the overall throughput to the transmit
power. For this utility function, we have shown that there exists a
unique Nash equilibrium for the proposed game, but, due to the frequency
selective multipath, this equilibrium is achieved at a different
output SINR for each user, depending on the channel realization and
the kind of Rake receiver used. Resorting to a large system analysis, we have
obtained a general characterization for the terms due to multiple access 
interference and self-interference, suitable for any kind of channel model 
and for different types of Rake receiver. Furthermore, explicit 
expressions for the utilities achieved at the
equilibrium have been derived for the case of an ARake receiver. 
It has also been shown that, under these hypotheses, the
noncooperative solution leads to a nearly SINR-balancing scenario.
In order to evaluate the efficiency of the Nash equilibrium, we
have studied an optimum cooperative solution for the case of ARake receivers, 
where the network seeks to maximize the sum of the users' utilities. It has 
been shown that the difference in performance between Nash and cooperative 
solutions is not significant for typical values of network parameters.

\appendices

\section{}\label{sec:proofs1_2}

\subsection{Proof of Theorem \ref{th:existence}}\label{pr:existence}
The following results are obtained from
\cite{saraydar2, debreu, fan, glicksberg}.

A Nash equilibrium exists
in the noncooperative game $\game = [\userset, \{\powerset[k]\},
\{\utility[k]{\powervect[]}\}]$ if, for all $k=1,\dots,\userno$:
\begin{enumerate}
\item $\powerset[k]$ is a nonempty, convex, and compact subset
  of some Euclidean space $\reals^\userno$; and
\item $\utility[k]{\powervect[]}$ is continuous in $\powervect[]$
  and quasi-concave in $\power[k]$.
\end{enumerate}

Each user has a strategy space that is defined by a minimum power
$\pmin[k]$ and a maximum power $\pmax[k]$, and all power values in
between. We also assume that $\pmax[k]\ge\pmin[k]$. Thus, the first
condition is satisfied. Moreover, since $\power[k]\ge 0$, it is
apparent from (\ref{eq:utility2}) and (\ref{eq:sinr}) that
$\utility[k]{\powervect[]}$ is continuous in $\powervect[]$.
To show that the utility function $\utility[k]{\powervect[]}$ is
quasi-concave in $\power[k]$ for all $k$ in the NPCG, it is sufficient
to prove that the local maximum of $\utility[k]{\powervect[]}$ is at
the same time a global maximum \cite{ponstein, roberts}.

For a differentiable function, the first-order necessary optimality
condition is given by
$\partial\utility[k]{\powervect[]}/\partial\power[k]=0$.
Recalling (\ref{eq:utility2}) and (\ref{eq:sinr}), the partial derivative
of $\utility[k]{\powervect[]}$ with respect to $\power[k]$ is
\be
  \label{eq:ut_der}
  \frac{\partial\utility[k]{{\powervect[]}}}{\partial\power[k]}=
  \frac{\infobits\rate[]}{\totalbits\power[k]^2}
  \left(\efficiencyFunctionPrime[{\SINR[k]}]\SINR[k]
  \left(1-\SINR[k]/\SIratio[k]\right)-\efficiencyFunction[{\SINR[k]}]\right),
\ee
where $\SIratio[k]$ is defined as in (\ref{eq:SIratio}) and
$\efficiencyFunctionPrime[{\SINR[k]}]=
d\efficiencyFunction[{\SINR[k]}]/d\SINR[k]$. For the sake of simplicity,
we do not explicitly show the dependence of $\SINR[k]$ on $\power[k]$.
Since $\power[k]\ge 0$ for the NPCG, we examine only positive real
numbers. Evaluating (\ref{eq:ut_der}) at $\power[k]=0$, we get
$\partial\utility[k]{\power[k], \powervect[-k]}/\partial\power[k]=0$.
Therefore, $\power[k]=0$ is a stationary point and the value of utility
at this point is $\utility[k]{\powervect[]}=0$. If we evaluate
utility in the $\varepsilon$-neighborhood of $\power[k]=0$, where
$\varepsilon$ is a small positive number, we notice that utility is
positive, which implies utility is increasing at $\power[k]=0$. Hence,
$\power[k]=0$ cannot be a local maximum. For nonzero values of the
transmit power, we examine the values of
$\SINRStar[k]=\SINR[k](\powerStar[k])$ such that
$\partial\utility[k]{\power[k], \powervect[-k]}/
\partial\power[k]|_{\power[k]=\powerStar[k]}=0$, thus satisfying the
first-order necessary optimality condition. In other words, we evaluate
$\SINRStar[k]$ such that
\be
  \label{eq:f_der2}
  \SINRStar[k]\left(1-\SINRStar[k]/\SIratio[k]\right)=
  \efficiencyFunction[{\SINRStar[k]}]/
  \efficiencyFunctionPrime[{\SINRStar[k]}],
\ee
as shown in (\ref{eq:f_der}).

We observe that the left-hand side of the above equation is a concave
parabola with its vertex in $\SINR[k]=\SIratio[k]/2>0$, and
$d\left(\SINR[k]\left(1-\SINR[k]/\SIratio[k]\right)\right)
/d\SINR[k]|_{\SINR[k]=0}=1$. The
right-hand side is an increasing function, with
$d\left(\efficiencyFunction[{\SINR[k]}]/
\efficiencyFunctionPrime[{\SINR[k]}]\right)
/d\SINR[k]|_{\SINR[k]=0}=1/i^\prime<1$ when 
$\efficiencyFunctionPrime[0]=0$, where
$i^\prime=\min\{i\in\naturals:d^i\efficiencyFunction[{\SINR[k]}]
/d\SINR[k]^i|_{\SINR[k]=0}\neq0\}$. Furthermore,
the equation is satisfied at $\SINR[k]=0$. Therefore, there is a single
value $\SINRStar[k]$ that satisfies (\ref{eq:f_der}) for $\SINR[k]>0$. The 
second-order partial derivative of the utility with respect to the power 
reveals that this point is a local maximum and therefore a global maximum. 
Hence, the utility function of user $k$ is quasi-concave in $\power[k]$ for 
all $k$. The same conclusion applies also if $\powerStar[k]>\pmax[k]$ 
for some $k$, even though $\SINRStar[k]$ cannot be achieved. In fact,
by applying the previous considerations to (\ref{eq:ut_der}), it is easy to 
verify that $\utility[k]{\power[k], \powervect[-k]}$ is strictly increasing 
in $\SINR[k]\in[0,\SINR[k](\powerStar[k])=\SINRStar[k])$, which in turn, from
(\ref{eq:sinr}), is strictly increasing in $\power[k]\in[0,\powerStar[k])$. 
Since $\power[k]\in[\pmin[k],\pmax[k]]$, which is a subset of 
$[0,\powerStar[k])$, $\utility[k]{\power[k]=\pmax[k], \powervect[-k]}$ 
represents both the local and the global maximum of the utility function.
\hfill$\blacksquare$

\begin{lemma}\label{lm:gamma_upper_bound}
  The solution $\SINRStar[k]$ of (\ref{eq:f_der}) satisfies the condition
  \be
    \label{eq:gamma_upper_bound}
    0\le\SINRStar[k]<\SIratio[k].
  \ee
\end{lemma}
\begin{proof}
  As $\efficiencyFunction[{\SINRStar[k]}]$ is an increasing function
  of $\SINRStar[k]$, $\efficiencyFunctionPrime[{\SINRStar[k]}]\ge 0$ for every
  $\SINRStar[k]$. Since the existence of the solution is ensured by
  Theorem \ref{th:existence} and $\SINRStar[k]$ and
  $\efficiencyFunction[{\SINRStar[k]}]$ are both greater than zero, the
  condition $(1-\SINRStar[k]/\SIratio[k])>0$ must hold.
\end{proof}

\subsection{Proof of Prop. \ref{prop:bestResponse}}\label{pr:bestResponse}

Using Theorem \ref{th:existence}, for a given interference, the SINR 
$\SINRStar[k]$ corresponds to the transmit power $\powerStar[k]$ as in 
(\ref{eq:power*}). Since $\SINRStar[k]$ is the unique maximizer of the 
utility, the correspondence between the transmit power and the SINR must 
be studied. As can be verified, (\ref{eq:power*}) represents the equation
of a hyperbola passing through the origin, with the asymptotes parallel
to the Cartesian axes. In particular, the vertical asymptote is
$\SINRStar[k]=\SIratio[k]$. Therefore, using Lemma
\ref{lm:gamma_upper_bound} presented in App. \ref{pr:existence}, 
there exists a one-to-one correspondence
between the transmit power, $\powerStar[k]\in[0,+\infty)$, and the SINR, 
$\SINRStar[k]\in[0,\SIratio[k])$. Thus, the transmit power $\powerStar[k]$
is also unique. If $\powerStar[k]\notin\powerset[k]$
for some user $k$, since it is not a feasible point, then $\powerStar[k]$
cannot be the best response to a given $\powervect[-k]$. In this case, 
we observe that $\partial\utility[k]{\powervect[]}/\partial\power[k]\le0$ 
for any $\SINR[k]\le\SINRStar[k]$, and hence for any 
$\power[k]\le\powerStar[k]$. This implies that the utility function is 
increasing in that region. Since $\pmax[]$ is the largest power in the 
strategy space, it yields the highest utility among all 
$\power[k]\le\pmax[]$ and thus is the best response to $\powervect[-k]$.
\hfill$\blacksquare$

\subsection{Proof of Theorem \ref{th:uniqueness}}\label{pr:uniqueness}
  By Theorem \ref{th:existence}, we know that there exists an equilibrium
  in the NPCG. Let $\powervect[]$ denote the Nash equilibrium in the NPCG.
  By definition, the Nash equilibrium must satisfy
  $\powervect[]=\bestresponsevect[{\powervect[]}]$, where
  $\bestresponsevect[{\powervect[]}]=[\bestresponse[1]
  (\powervect[]),\dots,\bestresponse[\userno](\powervect[])]$. The fixed
  point $\powervect[]=\bestresponsevect[{\powervect[]}]$ is unique if the
  correspondence $\bestresponsevect[{\powervect[]}]$ is a \emph{standard}
  function \cite{yates}, i.e., if it satisfies the following properties:
  \begin{enumerate}
  \item positivity: $\bestresponsevect[{\powervect[]}]>0$;
  \item monotonicity: if $\powervect[]>\powervect[]^\prime$, then
    $\bestresponsevect[{\powervect[]}]>
    \bestresponsevect[{\powervect[]^\prime}]$;
  \item scalability: for all $\mu>1$,
    $\mu\bestresponsevect[{\powervect[]}]>
    \bestresponsevect[{\mu\powervect[]}]$.
  \end{enumerate}
  It is apparent that $\bestresponse[k](\powervect[])=
  \bestresponse[k](\powervect[-k])$. Taking into account (\ref{eq:prop1}) 
  and (\ref{eq:power*}), the first condition translates into 
  $\powerStar[k]>0$ for all $k\in\userset$. Using (\ref{eq:sinr}), 
  (\ref{eq:hMAI}) and (\ref{eq:gamma_upper_bound}), the proof is 
  straightforward. Recalling (\ref{eq:prop1}) and (\ref{eq:power*}), the 
  second and the third condition are also apparent, since 
  $\powervect[-k]$ modifies only the numerator of (\ref{eq:power*}). 
  Therefore, since $\bestresponsevect[{\powervect[]}]$ is a standard
  function, the Nash equilibrium of the NPCG is unique.
  \hfill$\blacksquare$

\subsection {Proof of Prop. \ref{prop:requirement}}\label{pr:requirement}
  Based on Prop. \ref{prop:bestResponse}, when all users reach the
  Nash equilibrium, their transmit powers are
    \be
    \label{eq:powerNE}
    \powerStar[k]=
    \frac{\SINRStar[k]\left(\sum_{j\neq k}{\hMAI[kj]\powerStar[j]}+
      \sigma^2\right)}
     {\hSP[k]\left(1-\SINRStar[k]/\SIratio[k]\right)}.
  \ee
  Using Ass. \ref{ass:q} in (\ref{eq:powerNE}), it is
  straightforward to obtain:
  \be
    \label{eq:qNE}
    \powerTimesHsp\cdot
    \left[1-\SINRStar[k]\cdot
      \left(\SIratio[k]^{-1}+\MAIratio[k]^{-1}\right)\right]=
    \sigma^2\SINRStar[k]>0,
  \ee
  which implies
  $\SINRStar[k]\cdot\left(\SIratio[k]^{-1}+\MAIratio[k]^{-1}\right)<1$,
  proving necessity. It is also straightforward to show that, if each
  terminal $k$ uses transmit power $\powerStar[k]$ as in
  (\ref{eq:minimumPower}), all terminals will achieve the SINR
  requirement, finishing the proof of sufficiency. Finally, consider any
  other joint distribution of powers and channel realizations, and let
  $\powerTimesHsp^\prime=
  \inf_{k\in\userset}\left\{\hSP[k]\powerStar[k]\right\}$. Then, by
  exactly the same argument as was used in the proof of necessity,
  \be
    \label{eq:qPrime}
    \powerTimesHsp^\prime\ge
    \frac{\sigma^2\SINRStar[k]}
     {1-\SINRStar[k]\cdot\left(\SIratio[k]^{-1}+
       \MAIratio[k]^{-1}\right)}=
     \powerTimesHsp.
  \ee
  This means that assigning powers according to (\ref{eq:minimumPower})
  does indeed give the minimal power solution.
  \hfill$\blacksquare$

\section{}\label{sec:proofs_3_4}

In order to prove Theorems \ref{th:zeta} and \ref{th:gamma}, it is worth 
introducing the following results.
\begin{lemma}\label{Lemma2.7} \cite{baisil98}
Consider an $n$-dimensional vector $\bmr_n = [R_1,\dots, R_n]$ 
with independent and identically distributed (i.i.d.) 
standardized (complex) entries (i.e. $\Exop[R_i]=0$
and $\Exop[|R_i|^2]=1$, with $\Exop[\cdot]$ denoting expectation), 
and let $\bC_n$  be an $n\times n$ (complex) matrix
independent of $\bmr_n$. For any integer $p$,
\begin{align}
  \label{eq:Lemma2.7}
  \Exop[|\herm{\bmr_n} \bC_n \bmr_n - \Tr(\bC_n)|^p]&=
  K_p((\Exop[|R_1|^4]\Tr(\bC_n \herm{\bC_n}))^{p/2}\nonumber\\
  &+ \Exop[|R_1|^{2p}] \Tr(\bC_n \herm{\bC_n})^{p/2}),
\end{align}
where the constant $K_p$ does not depend either on $n$ or on $\bC_n$.
\end{lemma}

\begin{theorem}\label{th:quadraticForm}
  Consider an $n$-dimensional vector $\bmx_n = 
  \frac{1}{\sqrt{n}}[X_1,\dots, X_n]$ with
  i.i.d. standardized (complex) entries with finite eighth moment, and let 
  $\bC_n$  be an $n\times n$ (complex) matrix
  independent of $\bmx_n$ with uniformly bounded spectral radius for all $n$. 
  Under these hypotheses, 
  \be
    \label{eq:quadraticForm} 
    \herm{\bmx_n} \bC_n \bmx_n \as\tfrac{1}{n}\Tr(\bC_n).
  \ee
\end{theorem}
\begin{proof}
  Using Lemma \ref{Lemma2.7} and Markov's inequality,
  \begin{align}
    \Pr[|\herm{\bmx_n} \bC_n \bmx_n\!-\!\tfrac{1}{n}\!\Tr(\bC_n)|>
      \epsilon] &\le \frac{\Exop[|\herm{\bmx_n} \bC_n \bmx_n\!-
        \!\frac{1}{n}\!\Tr(\bC_n)|^4]}{\epsilon^{4}}\nonumber\\
    &\le\kappa\cdot\frac{1}{n^2\cdot\epsilon^4},
  \end{align}
  where $\kappa<\infty$ is a constant value independent of $n$. Thus, 
  \be
    \sum_{n=1}^{\infty}{\Pr[|\herm{\bmx_n} \bC_n \bmx_n-\Tr(\bC_n)|^4>
        \epsilon]}<\infty.
  \ee 
  Using the Borel-Cantelli lemma \cite{billingsley}, the result 
  (\ref{eq:quadraticForm}) is straightforward.
\end{proof}

\begin{theorem}\label{th:outerProduct}
  Suppose $\bmx_n = [X_1,\dots, X_n]$ and $\bmy_n = [Y_1,\dots, Y_n]$
  are $n$-dimensional independent vectors with
  i.i.d. standardized (complex) entries with finite eighth moment, and 
  $\bC_n$ is an $n\times n$ matrix (complex) independent on $\bmx_n$ 
  and $\bmy_n$ with uniformly bounded spectral radius for all $n$. Then,
  \be
    \label{eq:outerProduct} 
    \herm{\bmx_n} \bC_n \bmy_n \as  0.
  \ee
\end{theorem}
\begin{proof}\label{pr:outerProduct}
  The proof can be obtained using the same steps as that 
  of Theorem \ref{th:quadraticForm}.
\end{proof}

\subsection{Proof of Theorem \ref{th:zeta}}
  To prove that $\gain\MAIratio[k]^{-1}$ converges a.s. to non-random limits, 
  we focus on the ratio 
\begin{align}
  \gain\frac{\hMAI[kj]}{\hSP[j]}&=
    {\frac{\vectornorm{\herm{\matrakecoeff[k]}\cdot\vecpathgain[j]}^2
  + \vectornorm{\herm{\matpathgain[j]}\cdot\vecrakecoeff[k]}^2
  + \left|\herm{\vecrakecoeff[k]}\cdot\vecpathgain[j]\right|^2}
  {\left(\herm{\vecrakecoeff[k]}\cdot\vecpathgain[k]\right)
    \cdot\left(\herm{\vecrakecoeff[j]}\cdot\vecpathgain[j]\right)}}\nonumber\\
    \label{eq:zeta2}
    &={\frac{\frac{1}{\pathno^2}\!
        \left[\vectornorm{\herm{\matrakecoeff[k]}\cdot\vecpathgain[j]}^2\!
        +\!\vectornorm{\herm{\matpathgain[j]}\cdot\vecrakecoeff[k]}^2\!
        +\!\left|\herm{\vecrakecoeff[k]}\cdot\vecpathgain[j]\right|^2\right]}
      {\frac{1}{\pathno}
        \left(\herm{\vecrakecoeff[k]}\cdot\vecpathgain[k]\right)
        \cdot\frac{1}{\pathno}\left(\herm{\vecrakecoeff[j]}\cdot
        \vecpathgain[j]\right)}}.
\end{align}
It is sufficient to show that both numerator and denominator of 
(\ref{eq:zeta2}) converge a.s. to a non-random limit. Let
\begin{align}
  \label{eq:alphaWhite}
  \vecW[k]&=\left(\MatpathProfile[k]\right)^{-1}\vecpathgain[k]
\intertext{and}
  \label{eq:betaVersusAlpha}
  \vecrakecoeff[k]&=\processingMatrix\vecpathgain[k]\nonumber\\
  &=\left(\processingMatrix\circ\MatpathProfile[k]\right)\vecW[k]\nonumber\\
  &=\MatrakeProfile[k]\vecW[k],
\end{align}
where $\MatpathProfile[k]$ and $\MatrakeProfile[k]$ are defined as 
in (\ref{eq:MatpathProfile}) and (\ref{eq:MatrakeProfile}), respectively; 
the operator $\circ$ denotes the Hadamard (element-wise) product; 
and the matrix $\processingMatrix$ is dependent on the type of
Rake receiver employed. Using (\ref{eq:alphaWhite}) and 
(\ref{eq:betaVersusAlpha}), by Theorem \ref{th:quadraticForm}, we obtain
\begin{align}
  \frac{1}{\pathno^2}\vectornorm{\herm{\matrakecoeff[k]}\cdot
    \vecpathgain[j]}^2 
  &\as \functionPhi[{\frac{1}{\pathno}\MatpathProfile[j] 
      \matrakecoeff[k]\herm{\matrakecoeff[k]} \MatpathProfile[j]}] \nonumber\\
  &=\lim_{\pathno\rightarrow\infty} {\frac{1}{\pathno^2}
    \sum_{i=1}^{\pathno}{\{\MatpathProfile[j]\}_i^2
    \sum_{l=i+1}^{\pathno}{(\rakecoeff[l]{k})^2}}} \nonumber\\
  &=\lim_{\pathno\rightarrow\infty} {\frac{1}{\pathno^2}
    \sum_{l=1}^{\pathno-1}{(\rakecoeff[l+1]{k})^2
    \sum_{m=1}^{l}\{\MatpathProfile[j]\}_{m}^2}}\nonumber\\
  &=\lim_{\pathno\rightarrow\infty} {\frac{1}{\pathno}
    \sum_{l=1}^{\pathno-1}{\chi_l}},
\end{align}
 where $\functionPhi[\cdot]$ is defined as in (\ref{eq:Phi}) and
 \be
   \chi_l=\frac{1}{\pathno}(\rakecoeff[l+1]{k})^2
     \sum_{m=1}^{l}{\{\MatpathProfile[j]\}_{m}^2}
 \ee
 are independent random variables, with  
 \begin{align}
   \Exop\left[\chi_l\right] &=
   \frac{1}{\pathno}{\{\MatrakeProfile[k]\}^2_{l+1}}
     \sum_{m=1}^{l}{\{\MatpathProfile[j]\}_{m}^2}
 \intertext{and} 
 \Varop\left[\chi_l\right] &= 
 \frac{1}{\pathno^2}{\Varop[(\rakecoeff[l+1]{k})^2]
   \left(\sum_{m=1}^{l}\{\MatpathProfile[j]\}_{m}^2\right)^2}\nonumber\\
 &\le \Varop[(\rakecoeff[l+1]{k})^2]\cdot 
 \left(\frac{\Tr((\MatpathProfile[j])^2)}{\pathno}\right)^2\nonumber\\
 &< \infty.
 \end{align}

Using the weak version of the law of large numbers for non-i.i.d. 
random variables, 
\begin{align}\label{eq:num1}
  \frac{1}{\pathno}\sum_{l=1}^{\pathno-1}{\chi_l} 
  &\as 
  \lim_{\pathno\rightarrow\infty} {\frac{1}{\pathno}
    \sum_{l=1}^{\pathno-1}{\Exop[\chi_l]}}\nonumber\\
  &=\lim_{\pathno\rightarrow\infty} 
      {\frac{1}{\pathno^2}\sum_{l=1}^{\pathno-1}{\{\MatrakeProfile[k]\}^2_{l+1}
     \sum_{m=1}^{l}{\{\MatpathProfile[j]\}_{m}^2}}}\nonumber\\
  &=\functionPhi[{{\MatpathProfile[j]\MatrakeC[k]
            \herm{{\MatrakeC[k]}}\MatpathProfile[j]}}],
\end{align}
where $\MatpathC[k]$ and $\MatrakeC[k]$ are defined as in 
(\ref{eq:MatpathC}) and (\ref{eq:MatrakeC}), respectively. 
Similar arguments yield
\be\label{eq:num2}
  \frac{1}{\pathno^2}\vectornorm{\herm{\matpathgain[j]}\cdot
    \vecrakecoeff[k]}^2\as 
  \functionPhi[{{\MatrakeProfile[k]\MatpathC[j]
        \herm{{\MatpathC[j]}}\MatrakeProfile[k]}}].
\ee
Then applying Theorem \ref{th:outerProduct}, from 
(\ref{eq:betaVersusAlpha}) we obtain
\be\label{eq:num3}
  \frac{1}{\pathno}\herm{\vecrakecoeff[k]}\cdot\vecpathgain[j] \as 0,
\ee
since $\vecrakecoeff[k]$ is independent of $\vecpathgain[j]$. Analogously, 
using Theorem \ref{th:quadraticForm}, from (\ref{eq:alphaWhite}) we obtain
\be\label{eq:den}
  \frac{1}{\pathno}\herm{\vecrakecoeff[k]}\cdot\vecpathgain[k] \as 
  \functionPhi[{\MatpathProfile[k] \MatrakeProfile[k]}].
\ee
Using (\ref{eq:num1}) -- (\ref{eq:den}), the result (\ref{eq:zetaTh}) is 
straightforward.
\hfill$\blacksquare$

\subsection{Proof of Theorem \ref{th:gamma}}
In order to prove that $\gain/\SIratio[k]$ converges a.s. to a 
non-random limit, it is sufficient to show that both the numerator and 
the denominator converge to non-random limits. Note that 
\begin{multline}
  \vectornorm{\matCoeffHsi\cdot
      \left(\herm{\matrakecoeff[k]}\cdot\vecpathgain[k]+
      \herm{\matpathgain[k]}\cdot\vecrakecoeff[k]\right)}^2=\\
  \shoveleft{\qquad=\sum_{i=1}^{\pathno-1}{\coeffHsi[i]^2}\cdot
    \left(\sum_{l=1}^{i}{\pathgain[l]{k}\rakecoeff[\pathno+l-i]{k}
    + \sum_{l=1}^{i}{\rakecoeff[l]{k}}\pathgain[\pathno+l-i]{k}}
    \right)^2}\\
  \shoveleft{\qquad=\sum_{i=1}^{\pathno-1}{\coeffHsi[i]^2}\cdot
    \left(\sum_{l=1}^{i}
         {\functionTheta[k]{l, \pathno+l-i}\cdot
           \coeffW[k]{l}\cdot\coeffW[k]{\pathno+l-i}}
    \right)^2},\\
\end{multline}
where $\functionTheta[k]{l, \pathno+l-i}$ is defined as in (\ref{eq:Theta}).

Following the same steps as in the proof of Theorem \ref{th:zeta}, after 
some algebraic manipulation, it can be proven that
\begin{multline}\label{eq:num}
  \frac{1}{\pathno^2}\vectornorm{\matCoeffHsi\cdot
      \left(\herm{\matrakecoeff[k]}\cdot\vecpathgain[k]+
      \herm{\matpathgain[k]}\cdot\vecrakecoeff[k]\right)}^2
  \as\\
  \lim_{\pathno\rightarrow\infty}{\sum_{i=1}^{\pathno-1}{\coeffHsi[i]^2}\cdot
      \sum_{l=1}^{i}{\frac{\functionThetaQuad[k]{l,\pathno+l-i}}{\pathno^2}}}.
\end{multline}
Using (\ref{eq:num}), in conjunction with (\ref{eq:den}), the result 
(\ref{eq:gammaTh}) is straightforward.
\hfill$\blacksquare$


\section*{Acknowledgement}
The authors would like to thank S. Betz for many helpful discussions and 
useful comments.

\end{document}